\documentclass[final,5p,times,twocolumn]{elsarticle}

\pdfoutput=1
\input{glyphtounicode}
\usepackage[english]{babel}
\usepackage[utf8]{inputenc}
\usepackage{cmap}
\usepackage{amssymb, amsmath}

\usepackage{courier}
\usepackage{mathptmx}

\usepackage[T1]{fontenc}
\usepackage{textcomp}
\usepackage{graphicx}
\usepackage{xcolor}
\usepackage{microtype}
\usepackage[section]{placeins}

\usepackage[hyphens]{url}
\usepackage[unicode,
	pdfauthor={François-Xavier Coudert; Alain H. Fuchs},
	pdftitle={Computational characterization and prediction of metal-organic framework properties},
	pdfkeywords={metal--organic framework; molecular simulation; computational chemistry; theoretical chemistry; quantum chemistry; molecular dynamics; Grand Canonical Monte Carlo; Density Functional Theory},
	pdfcreator={},
	pdfproducer={}
]{hyperref}
\usepackage{bookmark}

\newcommand\e[1]{\ensuremath{_{\text{#1}}}}
\newcommand\ex[1]{\ensuremath{^{\text{#1}}}}

\usepackage{fancyhdr}
\fancypagestyle{plain}{
\fancyhf{}

\fancyhead[C]{\href{http://dx.doi.org/10.1016/j.ccr.2015.08.001}{\textcolor{blue}{Published as: \emph{Coord. Chem. Rev.} \textbf{2015}, DOI: 10.1016/j.ccr.2015.08.001}}}
}
\journal{Coordination Chemistry Reviews}

\let\oldfinalMaketitle\finalMaketitle
\renewcommand{\finalMaketitle}{\oldfinalMaketitle\tableofcontents\vspace{\baselineskip}\hrule\vspace{1cm}}

\begin{document}

\begin{frontmatter}
\title{Computational characterization and prediction of metal--organic framework properties}
\author{Fran\c{c}ois-Xavier Coudert}
\ead{fx.coudert@chimie-paristech.fr}
\ead[url]{@fxcoudert on Twitter}
\ead[url]{http://coudert.name/}
\author{Alain H. Fuchs}
\address{PSL Research University, Chimie ParisTech -- CNRS, Institut de Recherche de Chimie Paris, 75005 Paris, France}

\begin{abstract}
In this introductory review, we give an overview of the computational chemistry methods commonly used in the field of metal--organic frameworks (MOFs), to describe or predict the structures themselves and characterize their various properties, either at the quantum chemical level or through classical molecular simulation. We discuss the methods for the prediction of crystal structures, geometrical properties and large-scale screening of hypothetical MOFs, as well as their thermal and mechanical properties. A separate section deals with the simulation of adsorption of fluids and fluid mixtures in MOFs.
\end{abstract}


\end{frontmatter}

\thispagestyle{plain}

\section{Introduction}

\begin{table*}[t!]\centering
\begin{tabular}{lll}
\textbf{year} & \textbf{topic} & \textbf{ref.} \\\hline
2015	& general modeling of MOFs (book) & \citep{Jiang_book} \\
2015	& quantum chemical characterization of MOFs, including catalysis	&	\citep{Odoh2015} \\
2014	& high-throughput computational screening & \citep{Colon2014} \\
2014	& first-principles force fields for guest molecules &	\citep{Fang2014} \\
2013	& gas separation &	\citep{Yang2013} \\
2012  & screening for adsorption and separation	& \citep{Watanabe2012} \\
2012	& methane, hydrogen, and acetylene storage	&	\citep{Getman2012} \\
2011	&	adsorption in flexible MOFs &	\citep{Cou2011} \\
2011	&	energy, environmental and pharmaceutical applications &	\citep{Jiang2011} \\
2011  & screening for separation applications & \citep{Krishna2011} \\
2009	& hydrogen storage &	\citep{Han2009} \\
2009	& adsorption &	\citep{Duren2009} \\
2008	& adsorption and transport & \citep{Keskin2009} \\
2007	& adsorption of small molecules &	\citep{Nagaoka2007} \\
\hline\end{tabular}
\caption{\label{tab:reviews}List of reviews published on computational characterization of metal--organic frameworks.}
\end{table*}

Since its emergence in the 1950s, molecular simulation has seen an ever-growing use in research in the fields of physical, chemical, and materials sciences, where it offers an additional dimension to the characterization and understanding of systems, complementary to experimental techniques and pen-and-paper theoretical models. The development of novel computational methodologies, together with the exponential increase in computational power available to researchers, have dramatically expanded the range of problems that can be addressed through modeling. This is true of resource-intensive calculations performed on high-performance computing (HPC) supercomputers, but it is also true of desktop workstations, and even now of laptops and mobile devices.\cite{MolSim_iOS, Feldt2012} Several of the simulation techniques of computational chemistry have now reached the status of being relatively ``routine'' calculations and are nowadays considered an integral part of the researcher's toolbox, just like experimental characterization techniques like X-ray diffraction and NMR spectroscopy. Among those, we can cite Density Functional Theory calculations and Grand Canonical Monte Carlo simulations. It is possible to become a user of these tools with relatively little training, relying either on commercial or academic software with user-friendly interfaces. However, as with any technique, one should always take great care in checking the validity of the tools for the system at hand, as well as in interpreting the results obtained.

Given the formidable research effort focused on metal--organic frameworks (MOFs) in the past decade, with more than 20,000 papers published (at a current rate of more than 5 MOF papers per day\cite{MOF_papers}), 15,000 structures on record at the Cambridge Crystallographic Data Centre, and over 170 review articles dedicated to this topic, the published literature on computational studies of MOFs is in itself abundant. Theoretical approaches are in many cases used, in combination with experimental characterization techniques, to study newly synthesized materials and understand their properties at the microscopic level. In this \emph{introductory review}, we give an overview of the computational chemistry methods commonly used in the field of MOFs, to describe the structures themselves and characterize (or predict) their various properties. It is by no means a systematic review of existing computational work on MOFs, of which there simply is too much to systematically enumerate here. Rather, we will try to give the reader an idea of what is possible in theoretical studies of MOFs, both at the quantum chemical level and through classical molecular simulations. For more details on specific areas of interest, we refer to existing reviews of computational MOF studies, which are listed in Table~\ref{tab:reviews}.\footnote{In particular, we do not address in this review the very specific topic of the catalytic activity of MOFs: on this, we refer the reader to the very recent review by Odoh et al.\cite{Odoh2015}} We finish the review by pointing out some of the open questions and challenges in the field.

\section{Structural properties}

\subsection{Crystal structure prediction}

In most cases, the structure of a newly synthesized metal--organic framework is determined experimentally, using single-crystal X-ray diffraction data when possible, or solving the structure from powder diffraction data if single crystals of sufficient size or quality cannot be obtained. In the latter case, it often happens that because of the molecular complexity of the material, a low symmetry, or a large unit cell, the structure solution is arduous or impossible to solve. Moreover, in other cases, there is a need for methodologies of true (or \emph{ab initio}) computational prediction of MOF crystal structures, without any input of experimental data. These structures can then either be used to guide the synthesis of novel materials, or to identify materials already synthesized but whose crystals structure has not been solved experimentally, by comparison of X-ray powder diffraction patterns.

In the past decades, computational crystal structure prediction has made giant strides in both the fields of molecular crystals\cite{Price2014} and inorganic materials (dense and porous alike).\cite{Woodley2008} A large variety of methods have been developed and perfect to exploring the configurational space of these materials, including algorithms based on the simulated annealing approach, genetic (or evolutionary) algorithm methods,\cite{Kirkpatrick1983, Pannetier1990} and molecular packing approaches.\cite{Holden1993} There have also been techniques developed that are targeted specifically at framework materials, including both extended inorganic solids (such as zeolites) and hybrid inorganic--organic frameworks. Indeed, framework materials present a specific challenge when it comes to structure prediction, featuring both strong directional bonds and weaker dispersive interactions. We briefly summarize here the crystal structure prediction commonly used for metal--organic frameworks, and refer the reader to existing reviews\cite{Caroline_chapter, MellotDraznieks2007, AspuruGuzik_book} for a more comprehensive treatment of the subject.

Metal--organic frameworks, like inorganic framework materials, can be regarded as composed of elementary or secondary building units (SBU) assembled together into three-dimensional networks.\cite{Eddaoudi2001, Tranchemontagne2009} This fact is exploited in the \textbf{Automated Assembly of Secondary Building Units  (AASBU)} method for structure prediction. First developed for the computational prediction of inorganic extended lattices,\cite{MellotDraznieks2000} the AASBU method uses predefined SBUs with tailored ``sticky-atom'' interactions potentials between them, allowing the coordination in corner-, edge- and face-sharing modes. From these elements, the SBUs auto-assemble into three-dimensional frameworks through series of simulated annealing and energy minimizations. This strategy is applicable to MOFs as well as inorganic frameworks, as was demonstrated by the ``prediction'' of experimentally-known structures, including HKUST-1 and MOF-5, as well as novel hypothetical frameworks.\cite{MellotDraznieks2004} The AASBU method, which directly simulates \emph{in silico} the assembly of MOFs, is rather expensive computationally, as it requires series of minimizations and simulated annealings in order to converge structures.

\begin{figure*}[t]\centering
\includegraphics[width=0.8\linewidth]{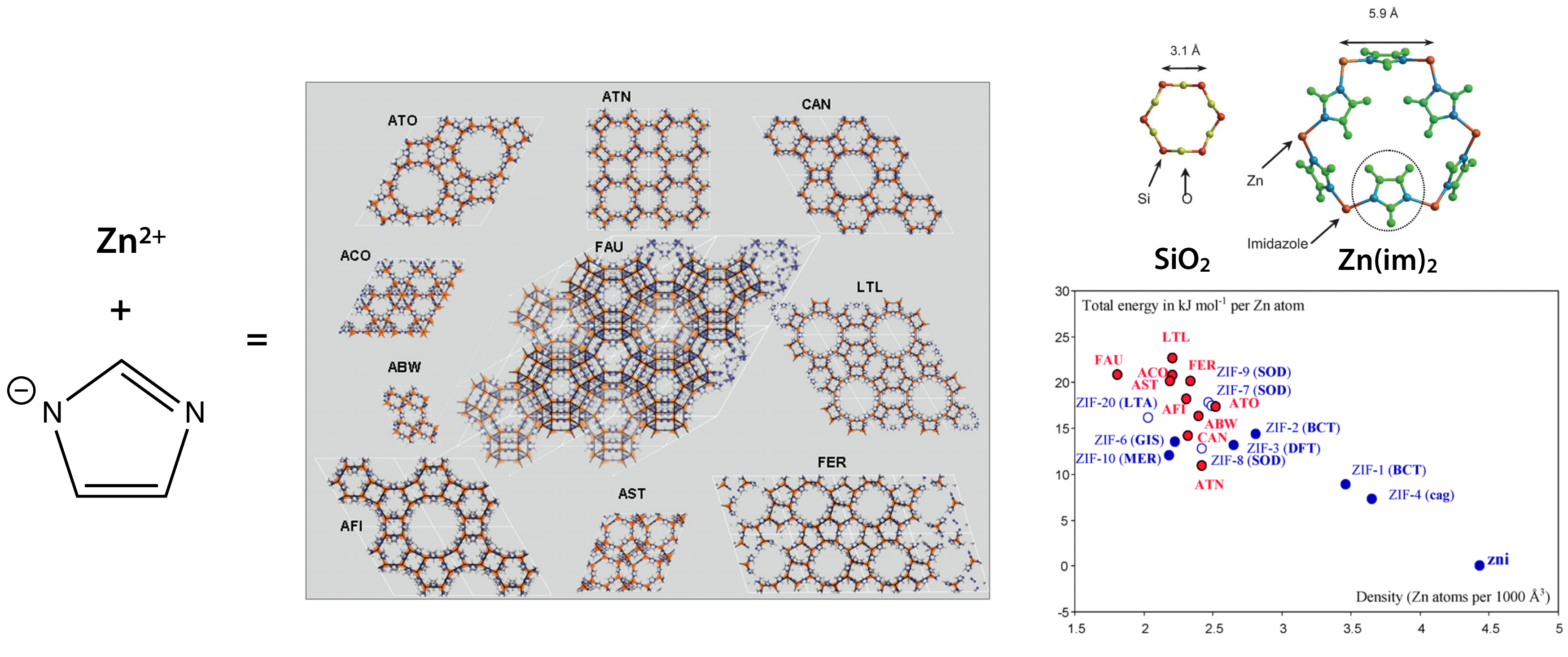}
\caption{\label{fig:ZIFs}Experimental and hypothetical zeolitic imidazolate frameworks (ZIF) generated by decoration of zeolite topologies by Lewis et al.,\cite{Lewis2009} along with a plot of their relative energy (compared to dense ZIF of \emph{zni} topology) versus density. Adapted from ref.~\citep{Lewis2009} with permission from The Royal Society of Chemistry.}
\end{figure*}

An alternative method is the direct construction of frameworks from SBUs by \textbf{direct enumeration of possible topologies} compatible with the ``building blocks'' available. In this top-down approach, sometimes called the ``decoration strategy'', the potentially infinite space of polymorphs of a given composition is explored by decorating a list of mathematical nets with the chosen SBUs, which act as vertices and edges of the net.\cite{Ockwig2005, OKeeffe2012} The nets themselves can be enumerated systematically from mathematical algorithms, given certain constraints on their features and complexity.\cite{Foster2003} They  can also be taken from a list of experimentally known structures, such as the database of known zeolitic structures (for zeolitic nets)\cite{IZA} or the broader Reticular Chemistry Structure Resource (RCSR) database.\cite{RCSR, OKeeffe2008} This approach has been used on a large variety of system, including: \begin{itemize} \item the identification of carboxylate-based MOFs with zeotype topologies, including the complex MIL-100 and MIL-101 structures;\cite{Ferey2004, Ferey2005} \item the enumeration of hypothetical (or ``not-yet-synthesized'', if one is optimistic!) Zeolitic Imidazolate Frameworks (ZIF),\cite{Baburin2008, Lewis2009} or more recently of porous zinc cyanide polymorphs;\cite{Trousselet2015} \item hypothetical covalent organic frameworks (COF).\cite{Bureekaew2013} \end{itemize}

The structures generated through this decoration process then need to be ``relaxed'' through energy minimization, relying on classical force field-based simulations or quantum chemistry calculations, in order to confirm their stability and evaluate their relative energy (or formation enthalpy). The latter is important in determining the experimental feasibility of structures: even if a given MOF structure is a local minimum in energy, if its relative energy is too high compared to other polymorphs, it cannot be considered a good target for possible synthesis. This process is exemplified on Figure~\ref{fig:ZIFs}, in the case of hypothetical ZIFs.\cite{Baburin2008, Lewis2009} Starting with a given composition (in this case, Zn$^{2+}$ cations and unsubstituted imidazolate anions), zeolitic nets from the IZA database\cite{IZA} are decorated with the metal centers (replacing the silicon atoms) and organic linkers (replacing the oxygen atoms). These ``ideal'' starting structures are energy-minimized through DFT calculations. The resulting structures can then be further studied, and their formation enthalpy (here, their energy relative to dense polymorph ZIF-\emph{zni}) give insight into their experimental feasibility. In the case of ZIFs, a general correlation between enthalpy and density is found, but the relatively small dispersion of the enthalpy values between hypothetical and experimental energies hints that the currently hypothetical frameworks should be attainable by solvothermal synthesis.

\begin{figure}[t]\centering
\includegraphics[width=\linewidth]{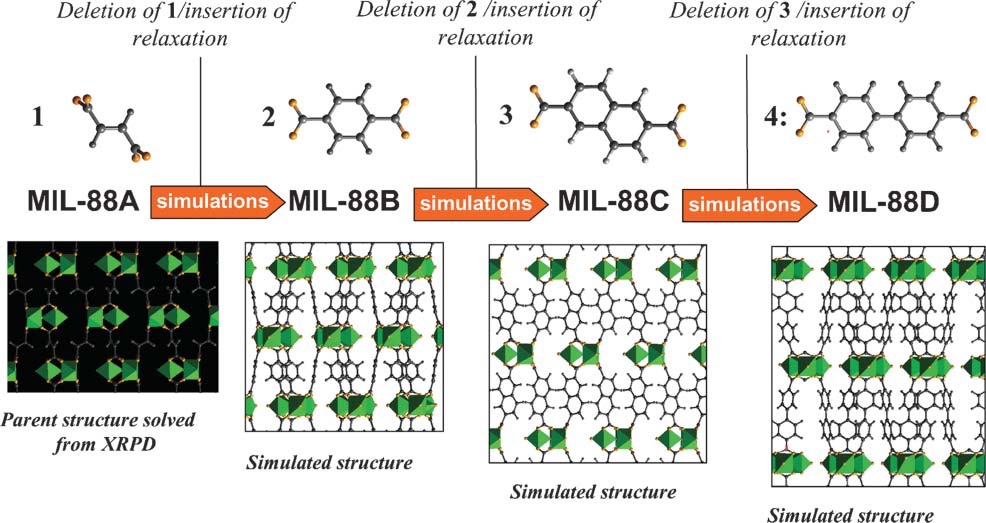}
\caption{\label{fig:MIL88}Depiction of the linker replacement strategy for computational prediction of isoreticular MOF structures, exemplified on the MIL-88 family. Reproduced from ref.~\citep{MellotDraznieks2007} with permission from The Royal Society of Chemistry.}
\end{figure}

Another property of metal--organic frameworks which can be leveraged for the computational prediction of new structures is the principle of \emph{isoreticularity}, i.e. the possibility of a family of MOFs to share the same topology and metal centers, with variations in their organic linkers both in terms of length and functionalization. This is best exemplified by the famous IRMOF series of structures,\cite{IRMOFs} which all possess the same topology as the original MOF-5 (aka IRMOF-1) structure.\cite{MOF5} Based on a given structure for a \emph{parent} MOF, it is possible to predict the structures of possible isoreticular analogues. Gradually replacing the original linker with longer organic molecules with the same coordination modes, and performing quantum chemical energy minimizations on each individual structure. This computational strategy is of interest in proposing structures of expanded versions of known structures. This was demonstrated in the case of the MIL-88 family of materials, a series of iron(III)-based and chromium(III)-based MOFs with linear dicarboxylic acid as linkers. There, the \textbf{ligand replacement strategy} was used for computational structure elucidation of the MIL-88B, MIL-88C, and MIL-88D compounds (see Figure~\ref{fig:MIL88}).\cite{Surble2006} Isoreticularity was also leveraged to design and screen new MOF-5 analogues based on commercially available organic linkers,\cite{Martin2013} and for the \emph{de novo} synthesis, after computational prediction, of ultrahigh surface area MOF NU-100.\cite{Farha2010}

\begin{figure*}[t]\centering
\includegraphics[width=0.8\linewidth]{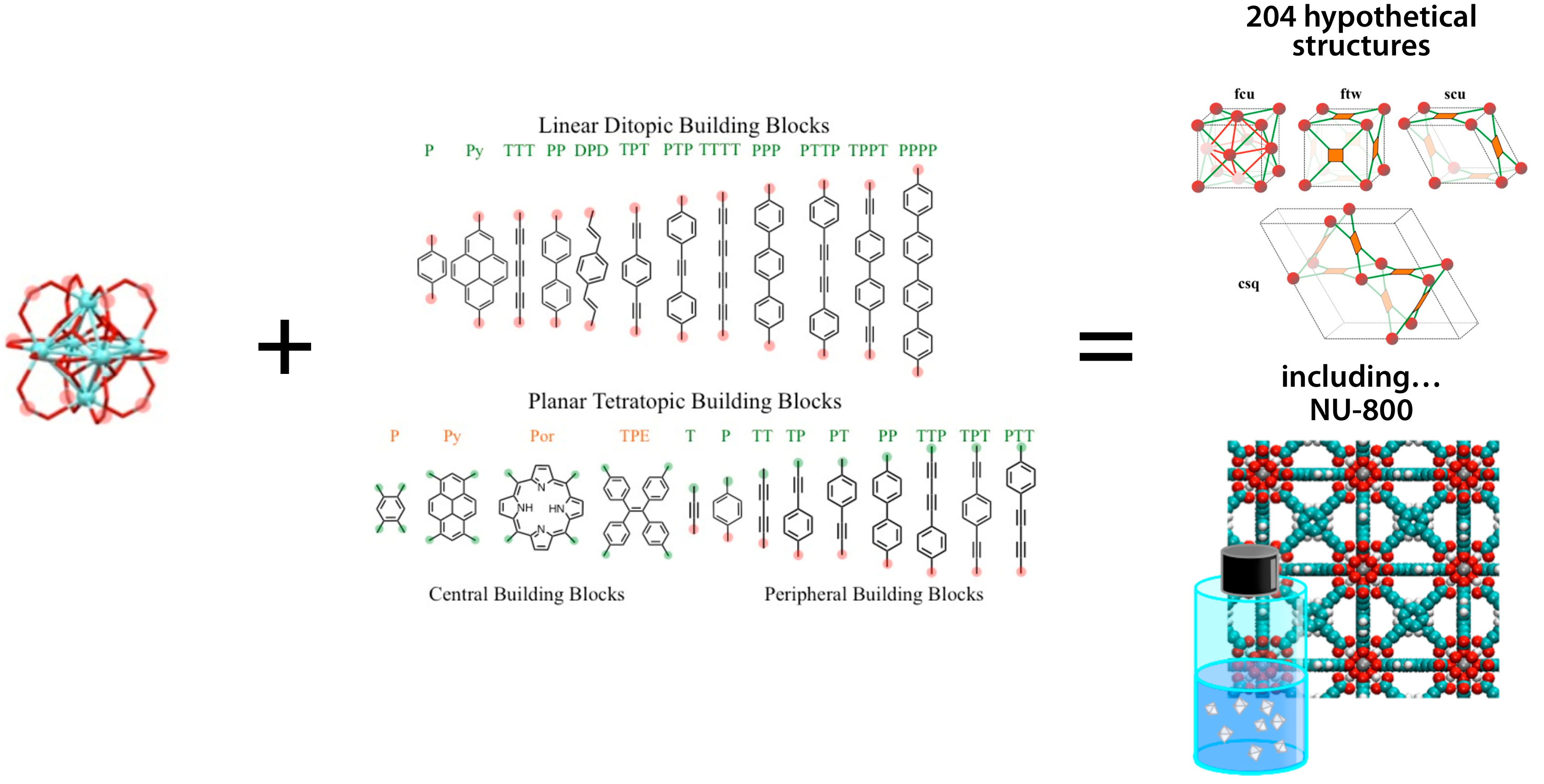}
\caption{\label{fig:GomezGualdron}Scheme of the reverse topological approach for crystal structure prediction, as implemented by Gomez-Gualdron et al.\cite{GomezGualdron2014} to generate 204 hypothetical (Zr\e{6}O\e{4})-based MOF structures. Adapted with permission from ref.~\citep{GomezGualdron2014}. Copyright 2014 American Chemical Society.}
\end{figure*}

As a final example, a combination of this linker replacement and functionalization approach with the ``decoration'' strategy highlighted above was used by Gomez-Gualdron et al.\cite{GomezGualdron2014} to generate 204 hypothetical MOF structures. In this \textbf{reverse topological approach},\cite{Bureekaew2013} the authors used as starting SBUs a zirconium-based metal center (Zr\e{6}O\e{4}), present in widely-studied material UiO-66(Zr), and 48 organic linkers (12 ditopic and 36 tetratopic building blocks). These building blocks were then combined in four possible topologies: \emph{fcu}, \emph{ftw}, \emph{scu} and \emph{csq} (see Figure~\ref{fig:GomezGualdron}). After computational identification of a top performer candidate for methane volumetric deliverable capacity, the predicted MOF NU-800 was synthesized and its high predicted gas uptake properties were confirmed by experimental high-pressure isotherm measurements over a large temperature and pressure range.

\subsection{Large-scale screening, enumeration of hypothetical structures\label{sec:screening}}

Although the different methods of crystal structure prediction described in the previous section each have different strengths and weaknesses, it is clear that they cannot necessarily scale to generate very large numbers of hypothetical structures. Yet, the constant search for new metal--organic frameworks has lead to an intense research effort to replace the current trial-and-error approach to MOF synthesis by enabling computationally-guided design of novel MOFs. There has thus been an effort, in the past few years, to enable larger-scale screening of potential MOF structures by generating larger number of hypothetical structures, and storing them into databases along with basic characterization information. This effort has been further spurred by being part of the broader Materials Genome Initiative, a 100 million dollar effort from the White House that aims to \emph{``discover, develop, and deploy new materials twice as fast''} as the current methods.\cite{MaterialsGenome1, MaterialsGenome2} In this section, we detail some of the recent milestones in the generation of large-scale databases of hypothetical materials and their screening for specific applications.\cite{Colon2014}

\begin{figure}[t]\centering
\includegraphics[width=\linewidth]{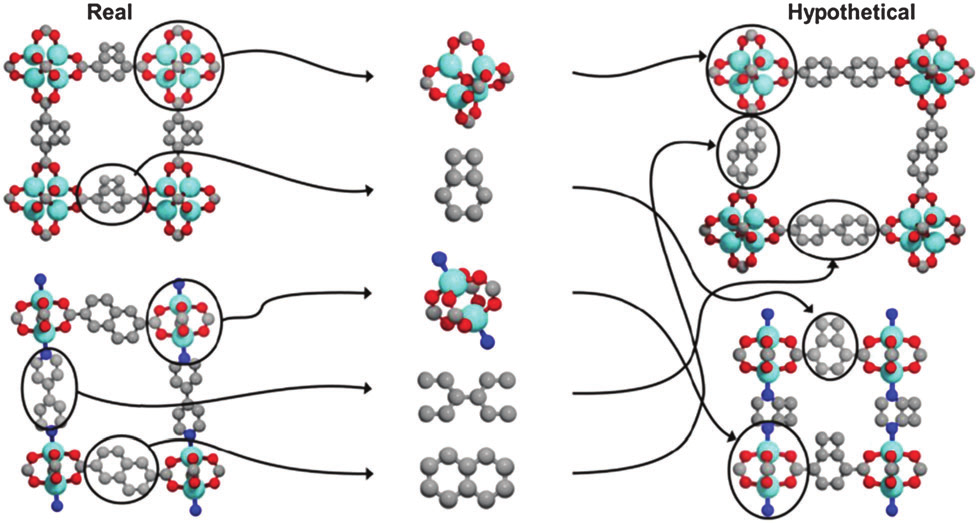}
\caption{\label{fig:Wilmer}Scheme of the bottom-up approach for large-scale generation of hypothetical MOFs by Wilmer et al.\cite{Wilmer2012}. Reproduced from ref.~\citep{Colon2014} with permission from The Royal Society of Chemistry.}
\end{figure}

Hypothetical databases of zeolite structures have been available for many years,\cite{Simperler2005, Foster_hypozeo, Rivin2006, Earl2006, Deem2009} containing up to two-million unique feasible structures. In the MOF world, the first large-size database of hypothetical material was the pioneering work of Wilmer et al. in 2012,\cite{Wilmer2012} a decisive step towards large-scale high-throughput screening of metal--organic frameworks. The generation procedure followed by Wilmer (depicted on Figure~\ref{fig:Wilmer}) is iterative, based on \textbf{step-wise combination of predefined building blocks relying on geometric rules of attachment}, pretty much like {LEGO\textregistered} bricks. All this information input into the generation procedure comes from the experimental literature: the building blocks are based on the reagents used in reported MOF syntheses, and the geometric rules for attaching the blocks together are determined by the crystallographic structures of known compounds using this particular coordination. This approach is more powerful than the decoration or linker replacement strategies, because it is entirely general and does not restrict to known topologies. It is also less computationally demanding than the assembly approaches such as AASBU, because it does not necessitate the minimization of energy (or scoring function). It thus scales to large number of structures, as Wilmer demonstrated by generating 137,953 hypothetical MOFs from a library of 120 building blocks, with the constraint that each MOF could contain only one type of metal node and one or two types of organic linkers, along with a single type of functional group.

A second example of the generation of hypothetical MOF structures comes from the family of zeolitic imidazolate frameworks (ZIF). Based on the extensive database of more than 300,000 hypothetical zeolite structures by Deem, \cite{Deem2009, Pophale2011} Lin et al. have used the decoration strategy to obtain a database of Zn(imidazolate)\e{2} ZIF structures, replacing the zeolites' silicon atoms by zinc cations and the bridging oxygen atoms by imidazolate linkers.\cite{Lin2012} In addition to these two hypothetical MOF databases, two publicly available databases of MOFs have been constructed from experimental structures. The first was gathered and published by Goldsmith et al.\cite{Goldsmith2013} Derived from the Cambridge Structural Database (CSD),\cite{Allen2002, CSD} the database contains \textbf{computation-ready MOF structures}, i.e. disorder-free structures from which solvent has been removed. The authors originally used it for the screening and selection of optimal hydrogen storage materials, as well as highlighting the theoretical limits of hydrogen storage in MOF materials.

\begin{figure*}[t]\centering
\includegraphics[width=0.55\linewidth]{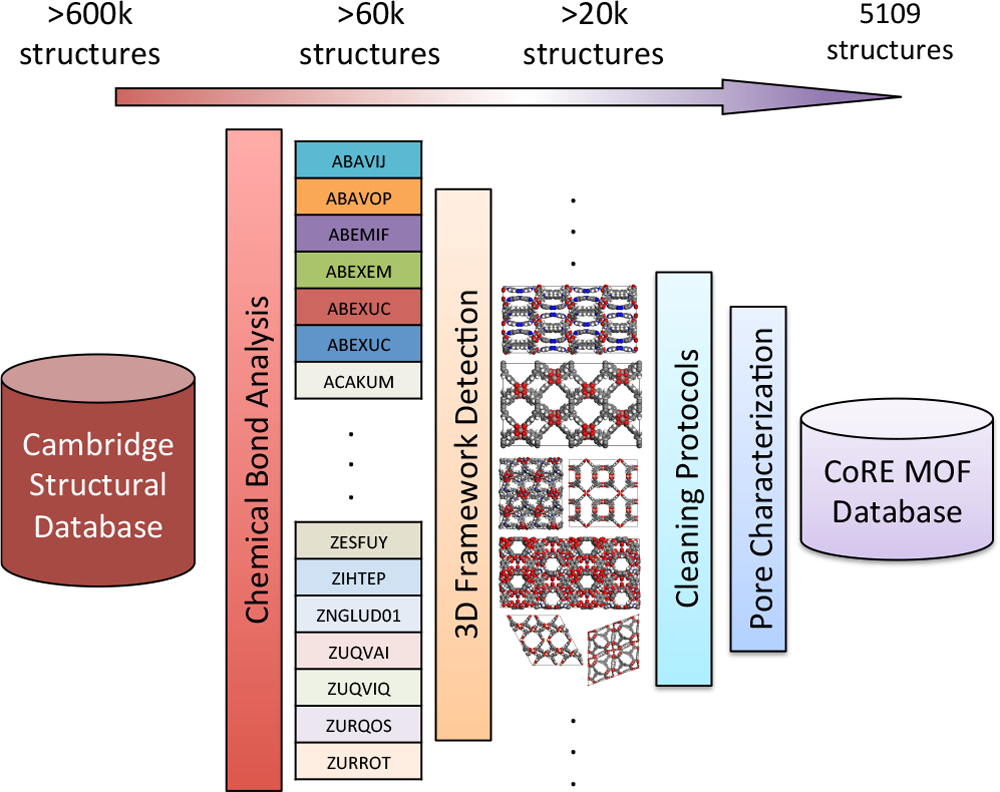}\hfill
\begin{minipage}[b]{0.4\linewidth}\centering
\includegraphics[width=\linewidth]{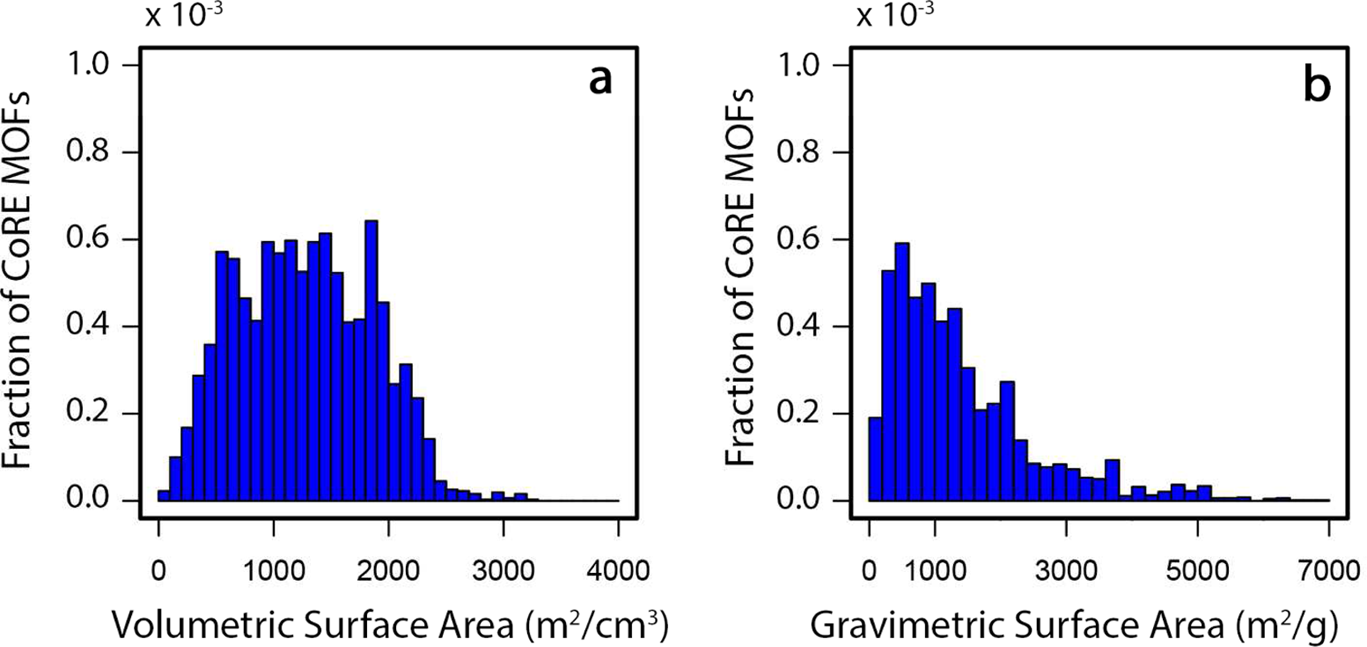}\\
\includegraphics[width=0.7\linewidth]{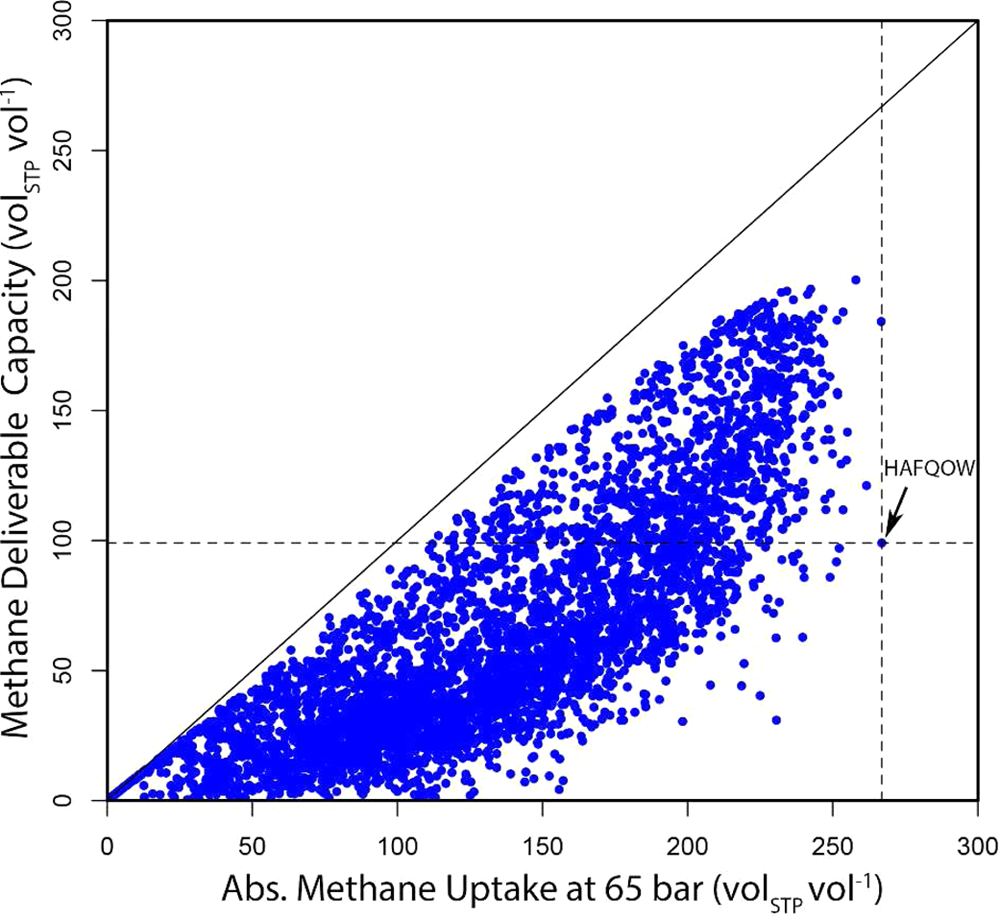}
\end{minipage}
\caption{\label{fig:CORE}Schematic illustration of the CoRE (Computation-Ready Experimental) MOF database construction.\cite{Chung2014} Adapted with permission from ref.~\citep{Chung2014}. Copyright 2014 American Chemical Society.}
\end{figure*}

A second database was built recently by Chung et al., with the aim of producing \emph{``a nearly comprehensive set of porous MOF structures that are derived directly from experimental data but are immediately suitable for molecular simulations or visualization''}.\cite{Chung2014} This database of computation-ready, experimental MOF structures (or CoRE MOF database) was produced from the CSD crystallographic structures through a multi-step procedure summarized in Figure~\ref{fig:CORE}: (i) selection of potential MOFs, based on chemical bond analysis (over 60,000 candidates after this step); (ii) elimination of 1-D coordination polymers and 2-D hydrogen bonded networks to retain only three-dimensional MOFs (over 20,000); (iii) removing partially occupied crystallographic positions, elimination of structures with disorder in the framework, identification of charge-balancing ions; (iv) solvent removal. The final CoRE MOF database contains over 4,700 hypothetical porous MOF structures. Its authors used it to investigate the structural properties of the CoRE MOFs that govern methane storage capacity in MOFs.

All the databases described above, whether of experimental and hypothetical structures, including zeolites, MOFs, porous polymer networks, and more, have been consolidated as part of The Materials Project and available online (upon registration) for browsing as well as systematic data mining through documented Application Programming Interfaces (APIs).\cite{MaterialsProject} Their use as a basis for large-scale high-throughput screening of porous materials has taken off in the past three years. Most of the screening studies proposed so far focus on identifying materials for adsorption-based applications, including separation,\cite{Sikora2012, Simon2015} capture,\cite{Lin2012} and storage\cite{Wilmer2012, Simon2015_EES, Colon2014_JPCC, Gomez2014} of strategic gases: hydrogen,\cite{Colon2014_JPCC, Gomez2014} methane,\cite{Wilmer2012, Martin2014, Simon2015_EES} carbon dioxide,\cite{Lin2012} noble gases,\cite{Sikora2012, Simon2015} etc. In these high-throughput screening approaches, the performance of materials are typically evaluated either by geometrical properties (pore size, pore volume, accessible surface area; see Section\ref{sec:geom}) or by evaluation of the host-guest interactions and adsorption properties (through Grand Canonical Monte Carlo simulations with classical force fields; see Section~\ref{sec:gcmc}). Large-scale high-throughput screening of metal--organic frameworks targeting other properties, or relying on other descriptors, has not developed so far.

\subsection{Geometrical properties: surface area, pore volume and pore size distribution\label{sec:geom}}

Once a MOF structure has been determined, either crystallographically or using a combination of diffraction experiments and quantum chemistry calculations, there are several so-called geometrical properties that can be calculated based solely on the unit cell parameters and atomic positions. Such geometrical calculations are straightforward to perform and computationally inexpensive, ranging from seconds to minutes on a desktop computer. They rely on a description of the material where each atom of the framework is a hard sphere, centered on crystallographic coordinates. The radii used for these hard spheres are the van der Waals radii of the atoms. Probe molecules, used in order to calculate guest accessibility, are also considered spherical, with a diameter equal to the kinetic diameter of the molecule.

\begin{figure}[t]\centering
\includegraphics[width=0.9\linewidth]{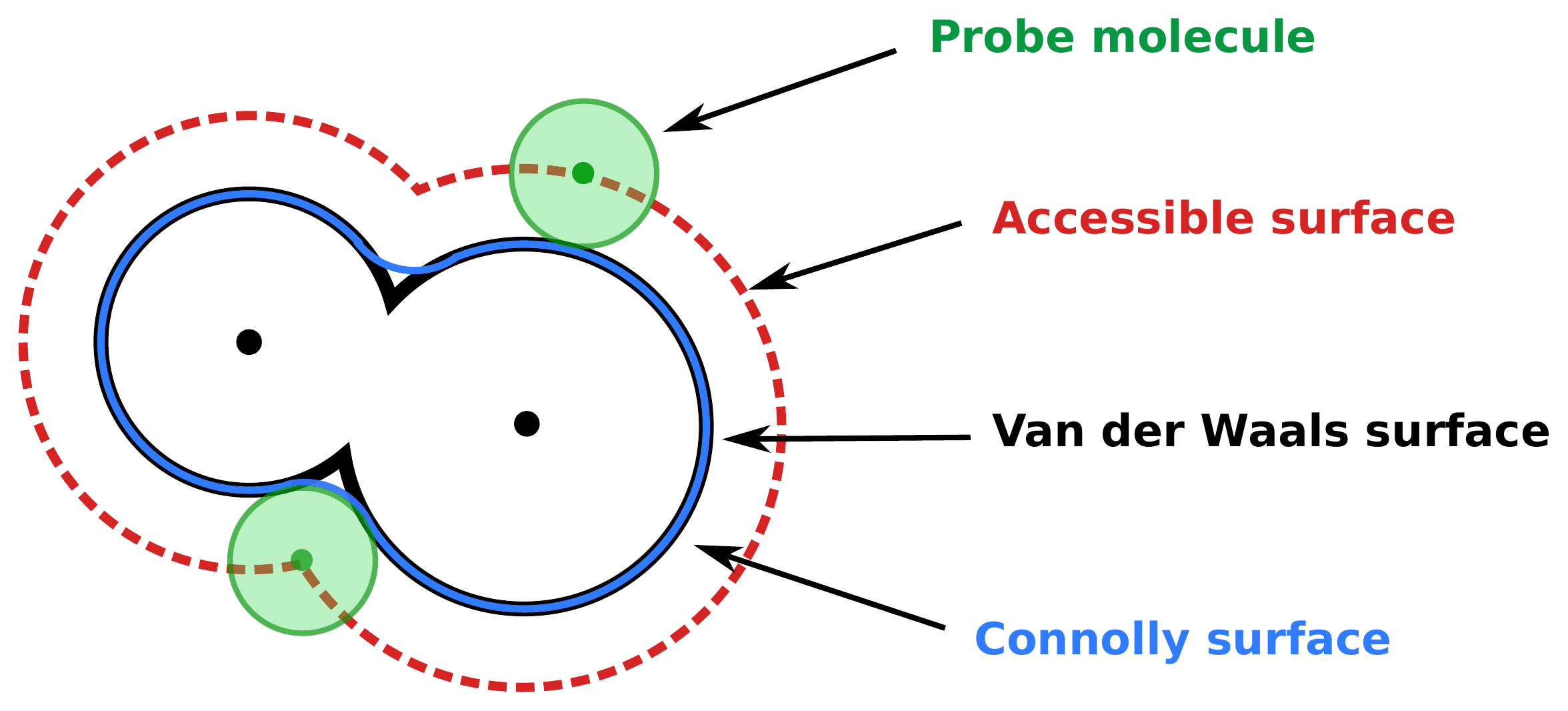}\\[5mm]
\includegraphics[width=0.7\linewidth]{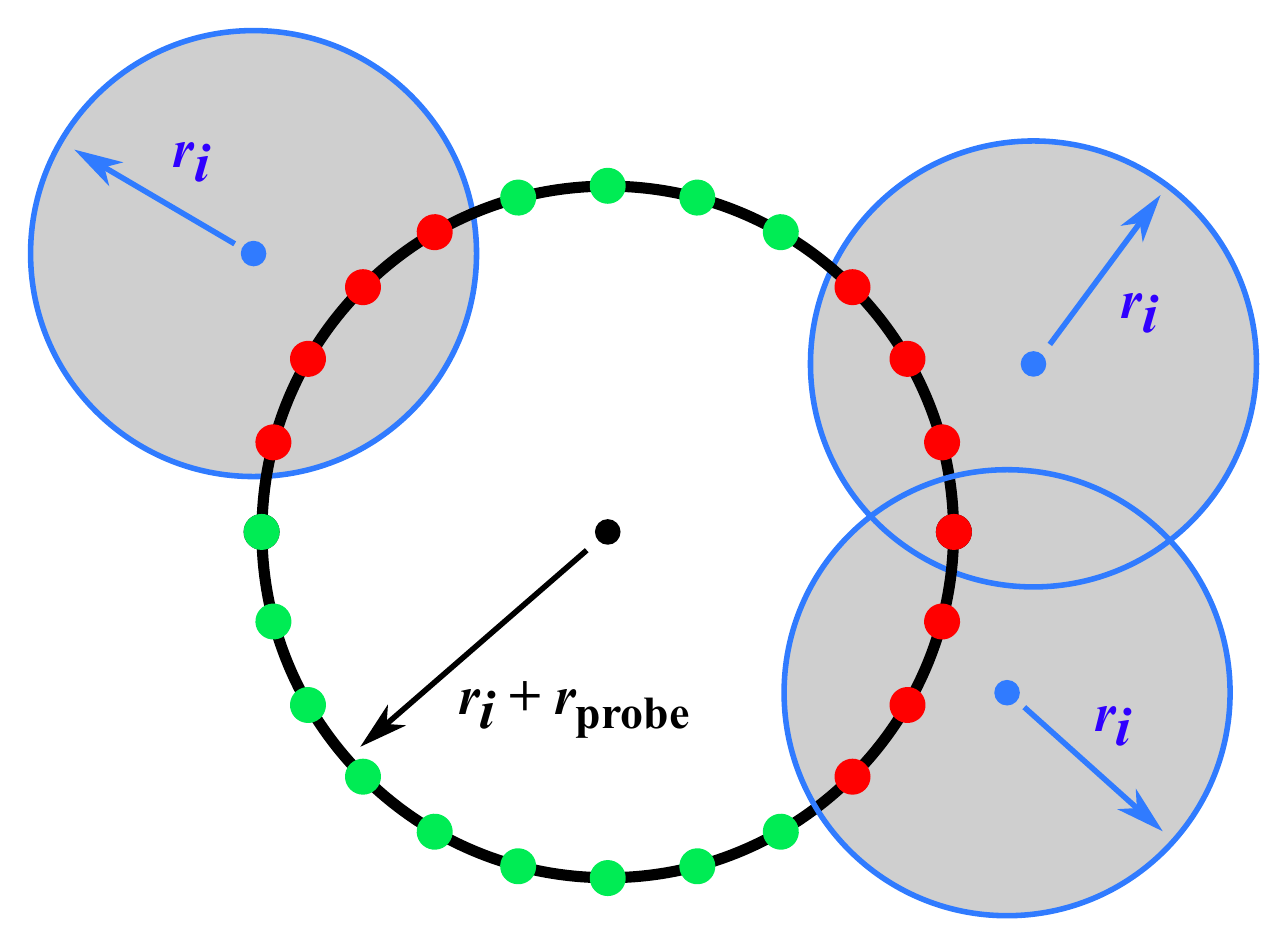}\\
\caption{\label{fig:surfaces}Upper panel: definition of the van der Waals surface (black line), Connolly surface (blue line), and accessible surface area (red dashed line). Lower panel: scheme of the calculation of accessible surface area contribution by each atom through sampling of its ``solvation sphere'': green points are accessible, red points are buried inside neighboring atoms and thus inaccessible.}
\end{figure}

The most common geometric characterization performed on microporous solids is that of their \textbf{accessible surface area} and \textbf{accessible pore volume}, representing the surface (resp. volume) accessible to guest molecules of a given size. Three possible surfaces can be used to measure this, as depicted in Figure~\ref{fig:surfaces}. The van der Waals surface is that defined by hard spheres centered on each atom, and does not depend on probe size. The accessible surface is that accessible to the center of mass of a probe of radius $r\e{p}$. Finally, the Connolly surface (also called solvent-excluded surface) delimits the space which is inaccessible to any part of the spherical probe; it is technically harder to compute. D\"uren et al. have shown that the accessible surface area is the appropriate surface area to characterize crystalline solids for adsorption applications.\cite{Duren2007} In particular, the accessible surface area calculated with a probe size corresponding to nitrogen ($r\e{p}=3.681$~{\AA}\cite{Bird}) can be directly compared with BET surfaces from experimental nitrogen isotherms\cite{Sing} (though experimental values might be lower because of blocked pores, or higher in presence of defects; see Section~\ref{sec:ads_exp}).

Sampling methods for the calculation of molecular surface are the most common numerical methods to evaluate the solvent-accessible surface of both molecules and periodic crystalline systems. This procedure is illustrated on Figure~\ref{fig:surfaces}'s lower panel: from a sample of points on each atom's ``solvation sphere'' (of radius $r_i+r\e{p}$), the proportion of points that are not buried inside neighboring atoms determines the atom's contribution to the total accessible surface area. This is called the Shrake-Rupley algorithm.\cite{Shrake1973} A similar sampling method can be adopted for the accessible pore volume: the pore space is sampled, e.g. on a regular mesh, and the number of mesh points falling within the pore volume is counted.\cite{PLATON_SOLV} Other sampling methods exist, however, for the determination of both accessible surface area and accessible pore volume, such as the use of ray casting\cite{Phillips_2010} and analytical calculations.\cite{Weiser1999, Klenin2011}

\begin{figure*}[t]\centering
\includegraphics[width=\linewidth]{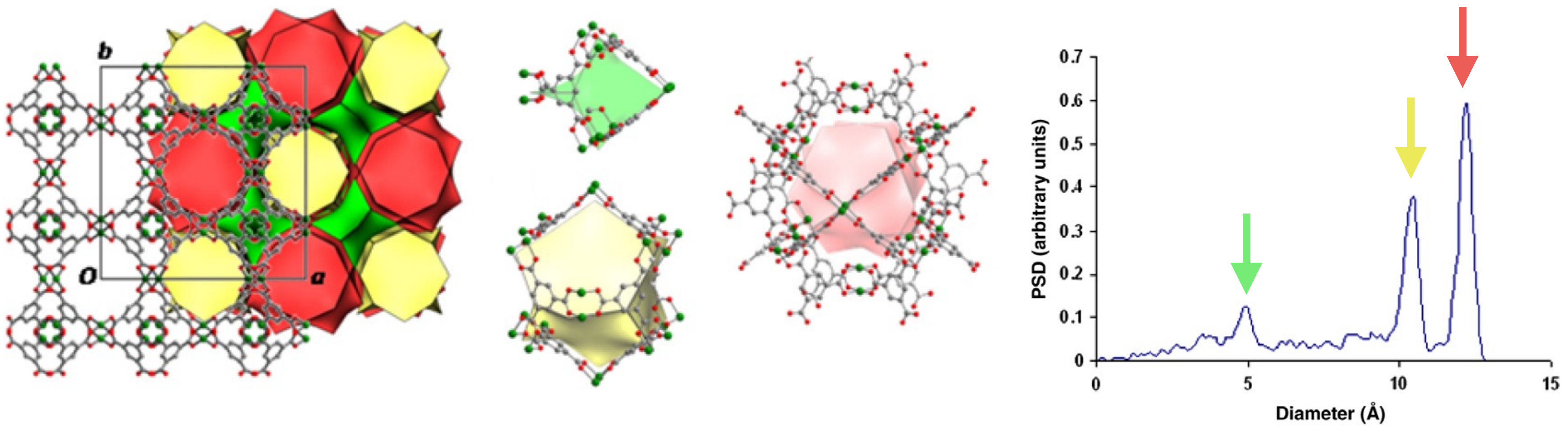}
\caption{\label{fig:PSD}Left: view on the structure of HKUST-1 (Cu\e{3}(btc)\e{2}), featuring small pores (green) and two different types of large pores (green and red). Right: pore size distribution (PSD) calculated from the crystal structure of HKUST-1, with peaks corresponding to the three types of pores. Adapted from ref.~\citep{Getzschmann2010}, with permission from Elsevier.}
\end{figure*}

Another tool at our disposal for geometric characterization of pore space is the \textbf{pore size distribution} (PSD), as illustrated on Figure~\ref{fig:PSD} in the case of metal--organic framework HKUST-1 (also known as Cu{3}(btc)\e{2}). Experimentally, pore size distributions can be obtained by numerical analysis of experimental low-temperature nitrogen or argon adsorption isotherms,\cite{Horvath1983, Landers2013} given the choice of a reference pore geometry (slit-like, cylindrical, spherical) and of an approximate chemical composition (though no kernels are available specifically for MOF materials). Thus, they provide a point of qualitative comparison between the geometric properties of the ideal crystal structure and those of the sample's pore space as observed through the adsorption process. The standard method to calculate geometrical pore size distributions was developed for Vycor glasses by Gelb and Gubbins,\cite{Gelb1999} and its computational efficiency was later improved by the same group.\cite{Bhattacharya2006} This method is entirely generic, applicable to nanoporous materials containing both micropores (below 2~nm) and mesopores (between 2~nm and 50~nm) and has been used with success in several MOF materials.\cite{Bae2009, Sarkisov2011}

\subsection{Advanced geometrical descriptors}

These methods described so far, however, do not account for the connectivity of the pore space determined by geometric means, nor for its accessibility to guest molecules along a diffusion path. For example, if a nanoporous structure presents a cavity linked to a single channel (side pocket), a guest may fit inside the cavity but not be able to enter in the first place if the channel is too narrow. It is therefore necessary to analyze the pore volume (and associated surface area) by breaking it down into \textbf{pathwise connected components}. Components of dimensionality equal to zero correspond to \textbf{isolated cages} or \textbf{unreachable side pockets}, which sorbate molecules cannot dynamically enter. Once identified, these need to be excluded from GCMC simulations of adsorption, in order to avoid insertion of molecules where it is not physically realistic.\cite{Krishna2010, Farrusseng2009} Components of higher dimensionality are one-dimensional \textbf{pore channels}, and 2D and 3D \textbf{pore networks}.

\begin{figure*}[t]\centering
\includegraphics[width=0.85\linewidth]{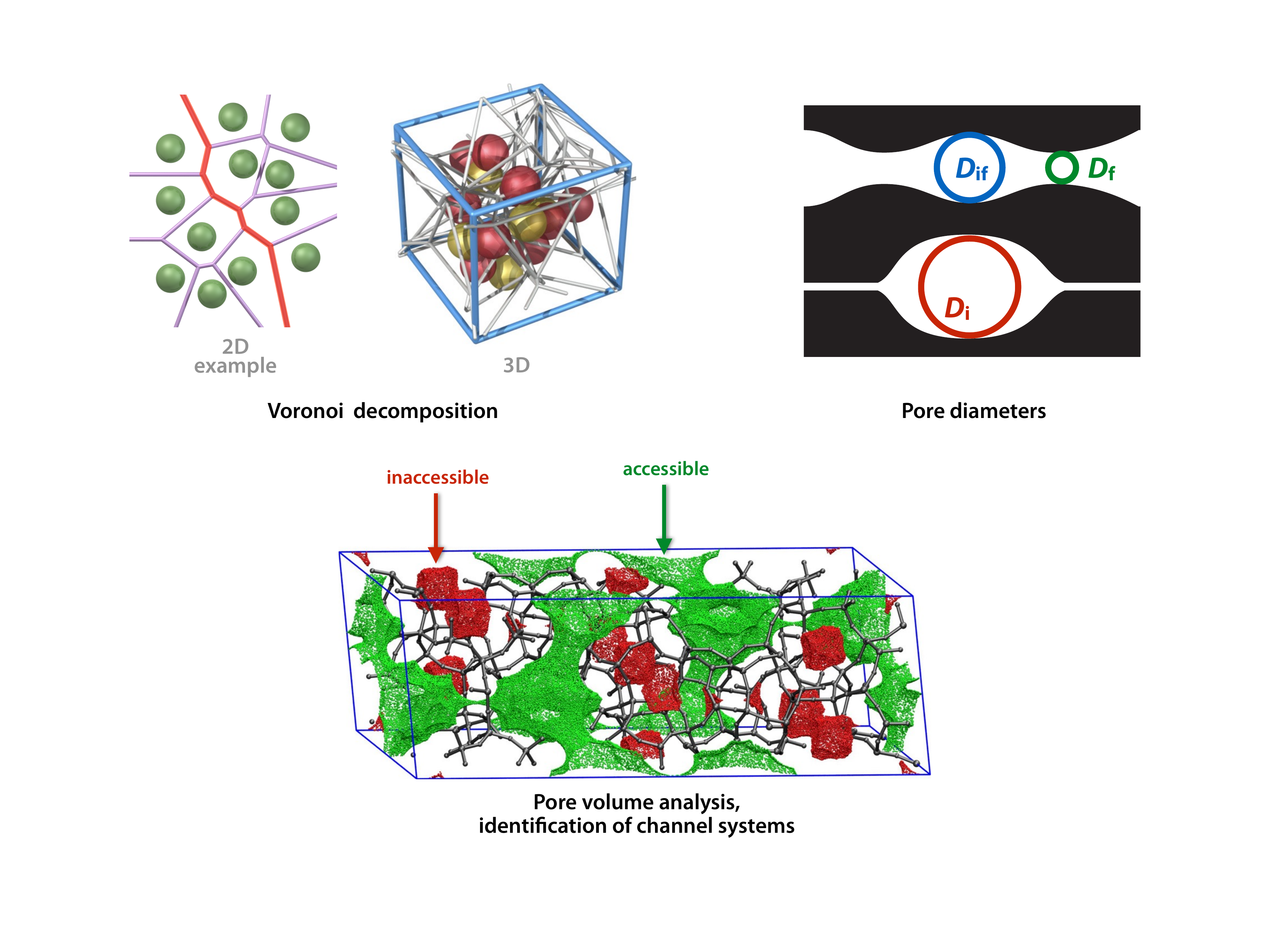}
\caption{\label{fig:diameters}Advanced geometrical characterization of nanoporous materials. Top left: depiction of the Voronoi decomposition of space in 2D and 3D. Top right: definition of the diameter of the largest included sphere ($D\e{i}$), diameter of the largest free sphere ($D\e{f}$), and largest included sphere along the free sphere path ($D\e{if}$). Bottom: decomposition of the pore volume of the DDR zeolitic framework (for a probe of radius {3.2~\AA}) between accessible and inaccessible volumes. Adapted from ref.~\citep{Willems2012}, with permission from Elsevier.}
\end{figure*}

The sampling methods described above for accessible surface and pore volume determination can be extended to further partition the porous network into connected components, typically by mesh-based propagation methods\cite{Haldoupis2010, Haldoupis2011, Haranczyk2010, Sarkisov2011} derived from the classical algorithms developed in the study of percolation theory.\cite{Hoshen1976} These grid-based methods are computationally expensive, in particular for high-accuracy determinations which require very fine mesh spacing. An alternative to the use of grid-based sampling to characterize pore accessibility exists, based on Voronoi decomposition (depicted on Figure~\ref{fig:diameters}),\cite{Blatov2003, Blatov2007} which for a given arrangement of atoms in a periodic domain provides a graph representation of the void space. The analysis of the Voronoi network and the accessibility of its nodes can then yield information into the components of the pore system, the dimensionality of the different channel systems, and the associated surface area and volume.\cite{Willems2012}

In addition, the analysis of the Voronoi network can yield other useful geometric descriptors for porous materials. Three quantities, in particular, are of particular relevance to the description of molecular adsorption and transport in porous materials (see Figure~\ref{fig:diameters}):\cite{Foster2006, Haldoupis2010, Willems2012} \begin{itemize} \item the \textbf{diameter of the largest included sphere}, $D\e{i}$, which reflects the size of the largest cavity within a porous material; \item the \textbf{diameter of the largest free sphere}, $D\e{f}$, representing the largest spherical probe that can diffuse fully through a structure (i.e. the size of the narrowest constriction in the channel system); \item the \textbf{largest included sphere along the free sphere path}, $D\e{if}$. \end{itemize} The definitions of these three diameters are illustrated on Figure~\ref{fig:diameters}. They can be used in order to better understand adsorption, separation and diffusion of guests in nanoporous materials,\cite{MatitoMartos2014} to quantify the similarities (and differences) between pore spaces of materials in a given family, as well as for high-throughput screening of structure databases.\cite{Willems2012, Colon2014}

From a practical point of view, the Voronoi-based analysis of nanoporous materials described above is implemented in the open-source Zeo++ software package.\cite{Zeo++, Willems2012, Martin2012} It allows in one single run the calculation of all geometric features of a given system. It is widely used for large-scale studies of zeolite and MOF databases and the screening of hypothetical novel structures.\cite{Martin2012}

Finally, while Voronoi-based descriptions of pore space appear as the most generic tool for systematic description of pore space geometries, it is worth noting that other approaches have been developed. We will cite here in particular the alternative method of First et al., which aims at fragmenting the pore space and representing as a set of geometrical blocks such as cylinders and spheres in order to identify portals, channels, cages, and their connectivity.\cite{First2011, First2013} Performing this analysis on a large number of available experimental crystal structures, First et al. built two online databases of nanoporous materials, ZEOMICS\cite{ZEOMICS} and MOFOMICS,\cite{MOFOMICS} aggregating quantitative information on the geometrical characteristics of zeolites and MOFs, respectively.

\begin{figure}[t]\centering
\includegraphics[width=\linewidth]{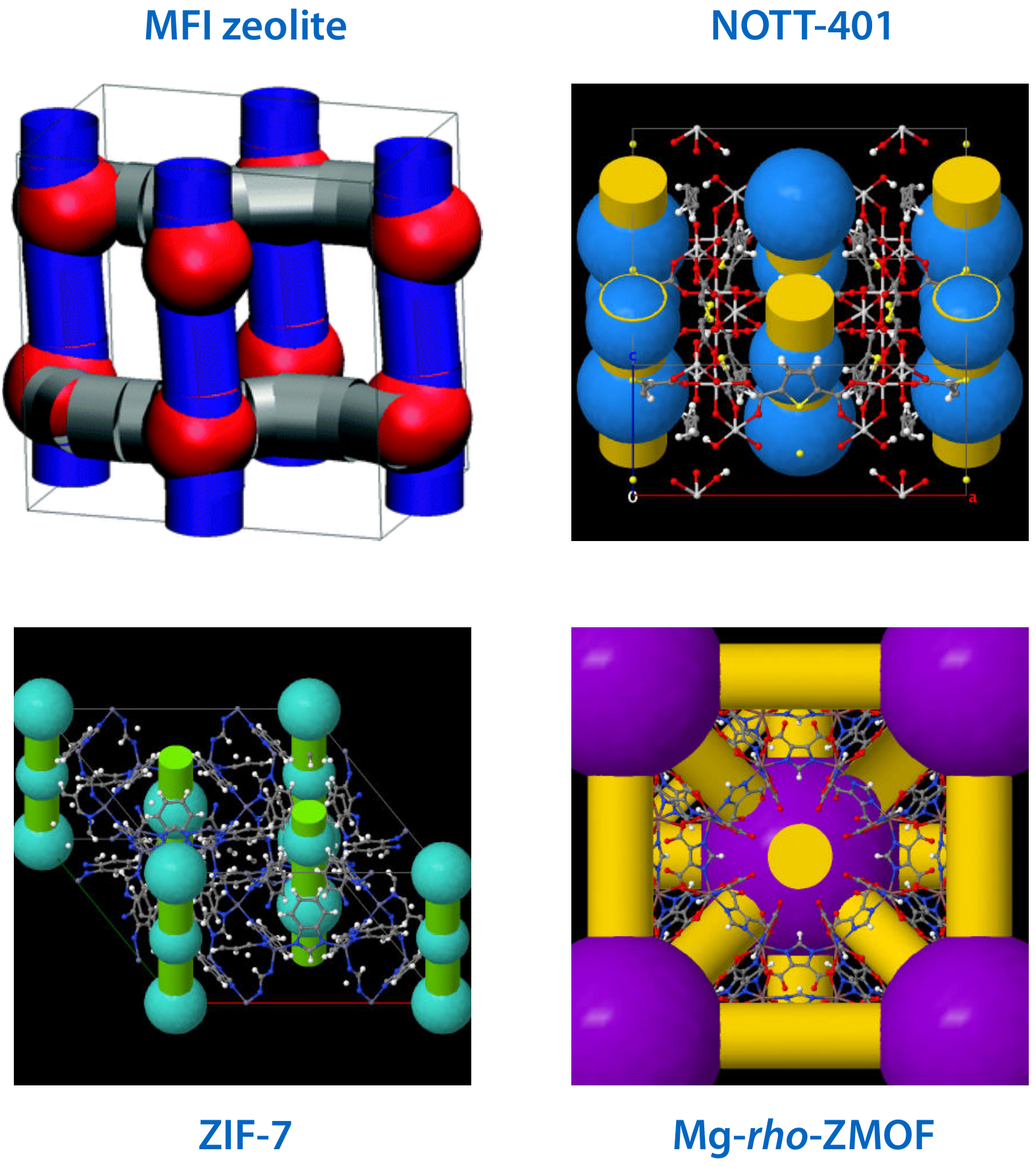}
\caption{\label{fig:MOFOMICS}Analysis of the pore systems of zeolite MFI and metal--organic frameworks NOTT-401,\cite{Ibarra2011} ZIF-7,\cite{Park2006} and Mg-\emph{rho}-ZMOF\cite{Nouar2009} by the ZEOMICS\cite{First2011} and MOFOMICS\cite{First2013} tools. Reproduced from ref.~\citep{First2011} with permission from The Royal Society of Chemistry, and the MOFOMICS website.\cite{MOFOMICS}}
\end{figure}

\subsection{Localization of extra-framework ions}

While most metal--organic frameworks possess a neutral framework, there is an important subclass of MOFs that feature ionic frameworks and charge-compensating extra-framework ions in their pores. These materials present some similarity to cationic zeolites, although in the case of MOFs both anionic frameworks (with extra-framework cations) and cationic frameworks (with extra-framework anions) are possible. In the past decade, a large number of ionic MOFs (sometimes called \emph{charged MOFs}) syntheses have been reported in the literature.\cite{Johnson2014} Perhaps the best-known materials in this family are the anionic zeolite-like metal--organic frameworks, or ZMOFs.\cite{Eddaoudi2015} $rho$-ZMOF and $sod$-ZMOF, synthesized by Eddaoudi and coworkers, are porous anionic MOFs with zeolitic topologies, whose overall neutrality is ensured by charge-balancing 1,3,4,6,7,8-hexahydro-2\emph{H}-pyrimido[1,2-\emph{a}]pyrimidine cations.\cite{Liu2006}

Both cationic and anionic metal--organic frameworks have been studied for potential applications. Due to the electric fields generated inside their nanopores by the presence of extra-frameworks ions, ionic MOFs show strong interactions with guest molecules, and specific interactions with polar guests in particular. This effect can in turn be leveraged for applications in gas adsorption and storage,\cite{Yang2009, Banerjee2011} separation,\cite{DeToni2012} molecular recognition,\cite{Chen2009} drug delivery,\cite{Bernini2014} ion exchange\cite{Tan2011CC} and catalysis.\cite{Fei2010a, Fei2010b, Zhang2012JACS} Since the extra-framework ions in charged MOFs can be exchanged, these materials offer a large versatility and tunability. However, the rational design of novel materials and their post-synthetic optimization requires a good understanding of their behavior at the microscopic level, and in particular of the \textbf{localization of their extra-framework cations} and the nature and strength of the \textbf{ion--guest interactions}.

The localization of extra-framework ions in the charged MOF structures is not always possible, in particular because the ions may be too delocalized or disordered. There, molecular simulation can play an important role, identifying possible binding sites for ions (i.e. local energy minima) and their physical characteristics: binding energy, mobility (i.e. spatial spread of the site). This can be achieved either at the level of quantum chemical calculations, or through molecular simulations based on classical force fields.  The latter, in addition to individual binding sites, can also help determine the distribution of cations between the various possible sites as a function of the experimental conditions: number and nature of cations, temperature, presence of solvent, etc. This is typically achieved through Monte Carlo simulations including large-scale displacement moves (or ``jumps''), in order to overcome the formidable energy barriers typically involved with movement of a cation from site to site. These simulations can also borrow methodologies from the very extensive scientific literature available on the topic of extra-framework cation localization in zeolites,\cite{Jaramillo2001, Maurin2004, Jeffroy2011} including the use of simulated annealing\cite{Maurin2001} or parallel tempering methods\cite{Beauvais2004, Earl2005} to reach equilibrium in reasonable time.

\begin{figure}[t]\centering
\includegraphics[width=0.9\linewidth]{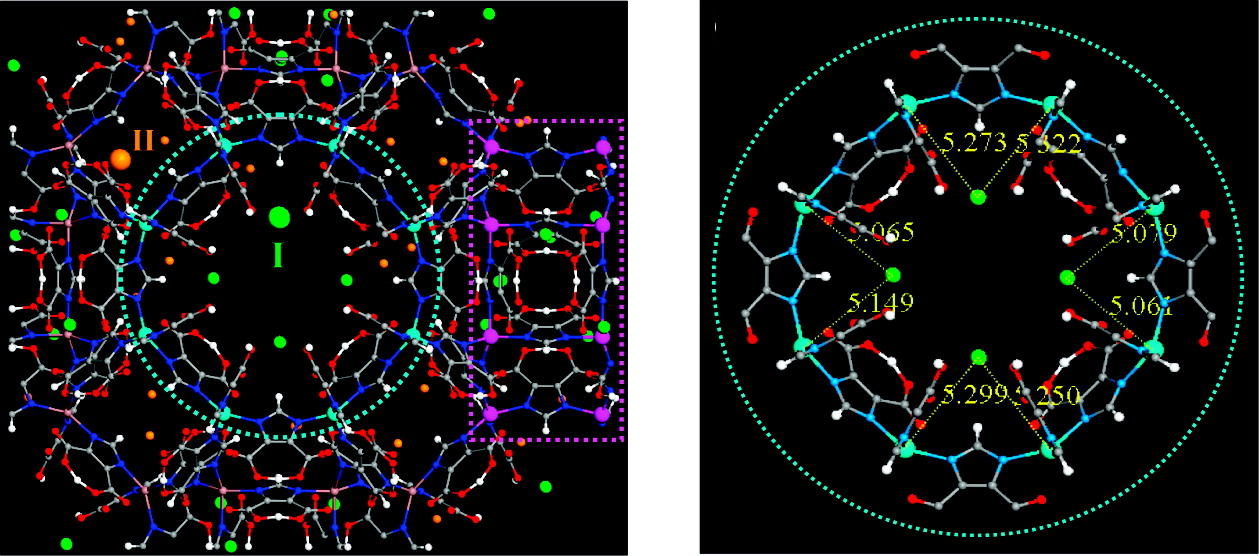}\\[3mm]
\includegraphics[width=0.9\linewidth]{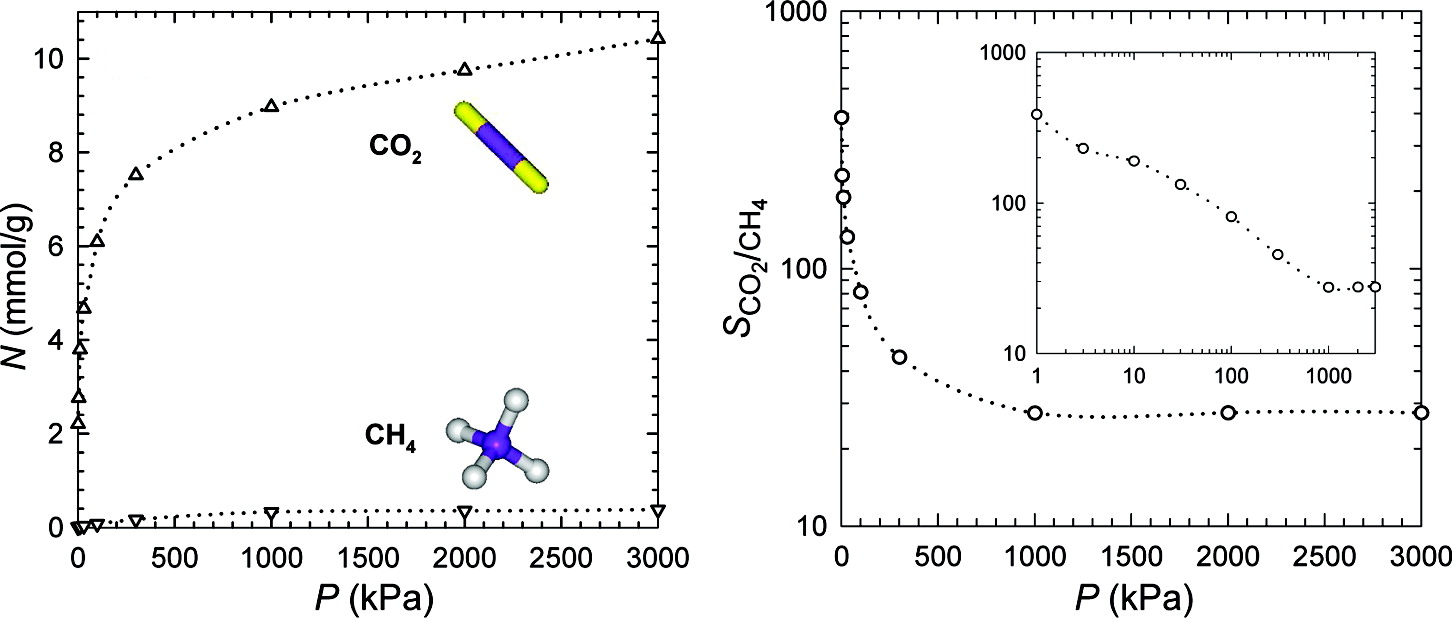}\\[3mm]
\includegraphics[width=0.9\linewidth]{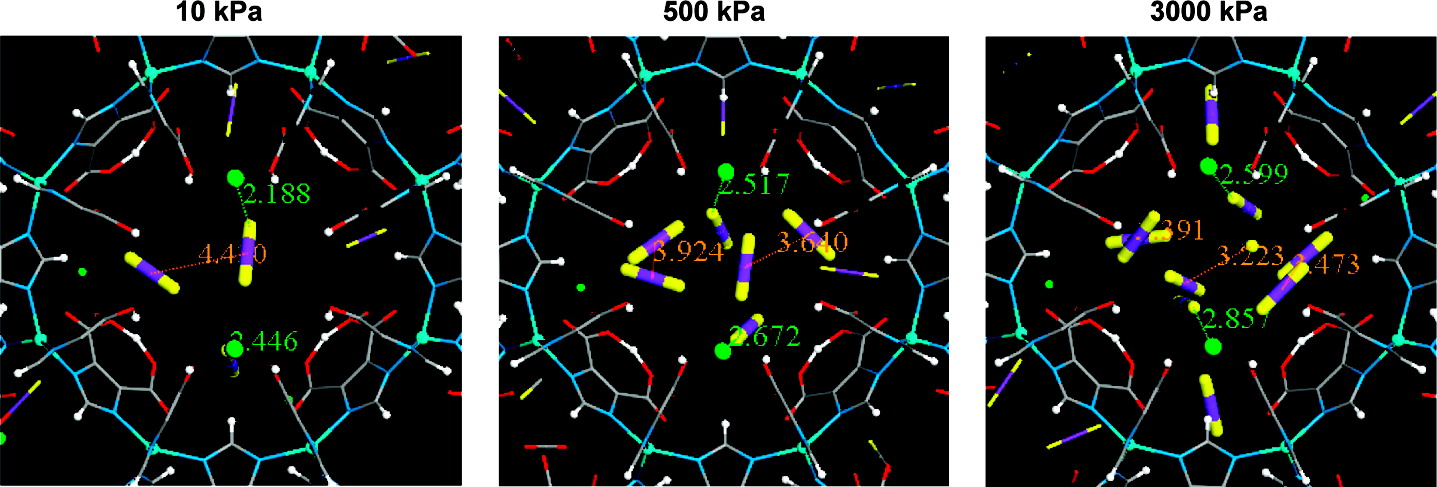}
\caption{\label{fig:ionic_MOF}Top: localization of the two cationic sites (I and II) for Na\ex{+} in the anionic \emph{rho}-ZMOF framework. Middle: Coadsorption isotherms and selectivity for an equimolar CO\e{2}/CH\e{4} mixture, calculated by Grand Canonical Monte Carlo. Bottom: Locations of CO\e{2} molecules adsorbed from the equimolar CO\e{2}/CH\e{4} mixture, at various pressures. Reproduced with permission from ref.~\citep{Babarao2009}. Copyright 2009 American Chemical Society.}
\end{figure}

Once the localization of extra-framework ions has been determined, it can then be used as input or basis for the study of adsorption properties of the ionic MOF.\cite{Jiang2009, Babarao2009Langmuir, Babarao2011, DeToni2012} These are methodologically similar to simulations of adsorption in neutral MOFs, except for the very large strength of the Coulombic ion--guest interactions. While simulations of adsorption are treated in detail in Section~\ref{sec:adsorption}, we give in Figure~\ref{fig:ionic_MOF} a rather typical example of results from molecular simulation in ionic MOFs, here in the case of the anionic \emph{rho}-ZMOF with extra-framework Na\ex{+} cations. In this study, Babarao et al.\cite{Babarao2009} identified two types of binding sites for Na\ex{+} ions in the anionic framework, with site~I in an eight-membered ring and site~II in the $\alpha$-cage. The authors showed that carbon dioxide is adsorbed predominantly over other gases, including methane, because of its strong electrostatic interactions with the charged framework and the presence of Na\ex{+} ions acting as additional adsorption sites.

\section{Physical properties}

\subsection{Mechanical properties\label{sec:mechanical}}

\begin{figure}[t]\centering
\includegraphics[width=0.6\linewidth]{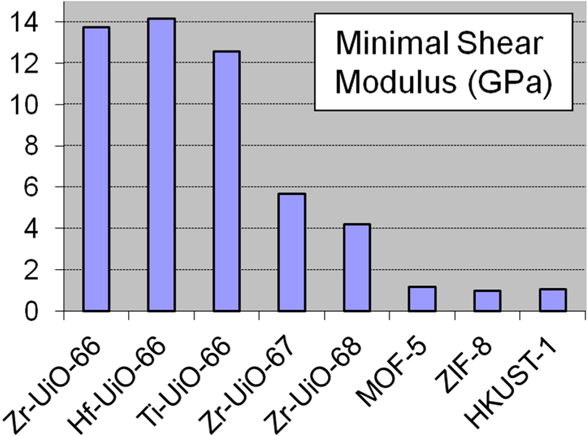}\\[4mm]
\includegraphics[width=\linewidth]{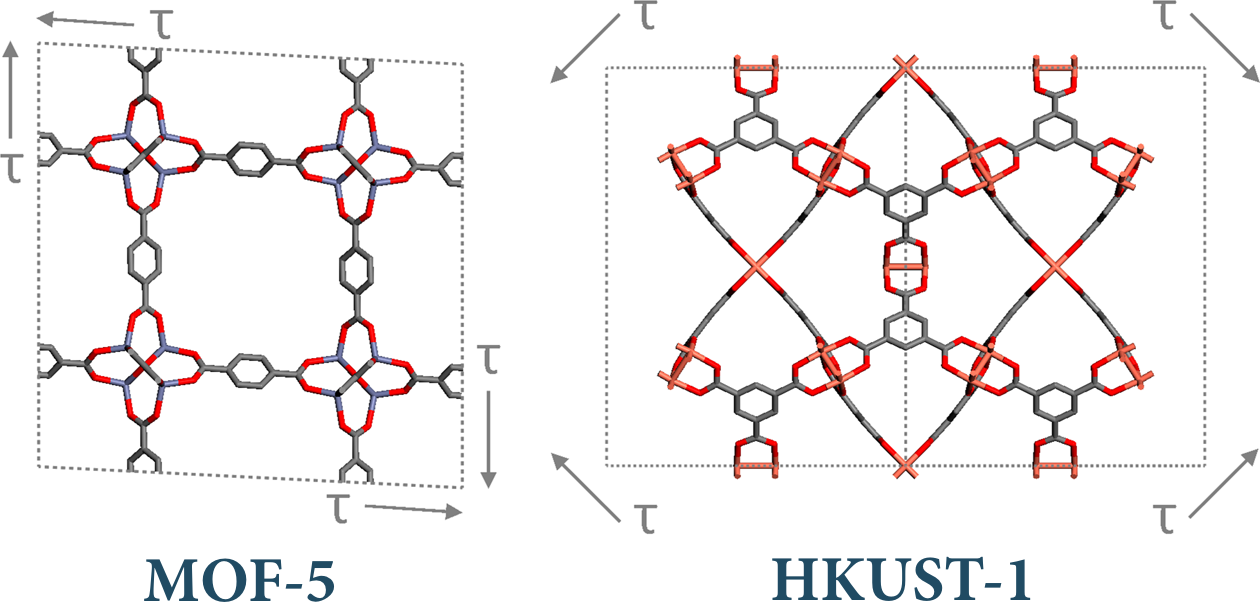}
\caption{\label{fig:Wu2013}Top: comparison of the minimal shear modulus of several metal--organic frameworks, obtained from DFT calculations. Bottom: representation of the soft shear modes of MOF-5 and HKUST-1. Reproduced with permission from ref.~\citep{Wu2013}. Copyright 2013 American Chemical Society.}
\end{figure}

Compared to structural characterization and adsorption properties, studies of the mechanical properties of metal--organic frameworks appeared relatively late in the literature, both experimental and computational. The first theoretical calculations of mechanical properties were predictions of the \textbf{bulk modulus} of MOF-5 (and several analogues) by DFT calculations:\cite{FuentesCabrera2005} Fuentes-Cabrera et al. calculated the energy vs. volume curves for each of these cubic materials, fully relaxing the atomic positions for each value of unit cell parameter $a$; then the curves were fitted by the Birch--Murnaghan equation of state.\cite{Birch1952} Later work performed calculations not only of bulk modulus, but also the individual \textbf{elastic constants} ($C_{11}$, $C_{12}$, and $C_{44}$) of the cubic MOF-5,\cite{Mattesini2006, Samanta2006, Zhou2006, Bahr2008} along with its average Young's and shear moduli. These calculations were again performed at the DFT level (either with LDA or GGA exchange--correlation functions). They employed either the fitting of quadratic ``energy vs. strain'' curves, or linear ``stress vs. strain'' curves, by applying small strains to the relaxed structure.

Since these seminal studies, other cubic materials have been studied by the same methods, including IRMOFs\cite{Kuc2007}, ZIF-8\cite{Tan2012}, UiO-66,\cite{Wu2013} etc. In addition to these zero~Kelvin quantum chemical calculations, other works have reported bulk moduli at finite temperature, through force field-based molecular dynamics simulations, e.g. for MOF-5\cite{Han2007} and HKUST-1.\cite{Tafipolsky2010} The main conclusion, from both experimental and computational work on elastic properties of MOFs,\cite{Tan2010, Bennett2010, Tan2011, Tan2012} is that MOFs are far \emph{softer} than inorganic nanoporous materials such as zeolites, i.e. they present much lower elastic moduli, though their detailed mechanical properties depend on both chemical composition and the framework's geometric properties (see Figure~\ref{fig:Wu2013}). In particular, many highly porous MOFs feature very low shear moduli (of the order of 1~GPa), implying relatively small mechanical stability. The UiO-66 family of materials is one of the exceptions, with shear modulus in the 12--14~GPa range, due to its strong Zr--O linkages and high degree of coordination.\cite{Wu2013}

\begin{figure*}[t]\centering
\includegraphics[width=0.9\linewidth]{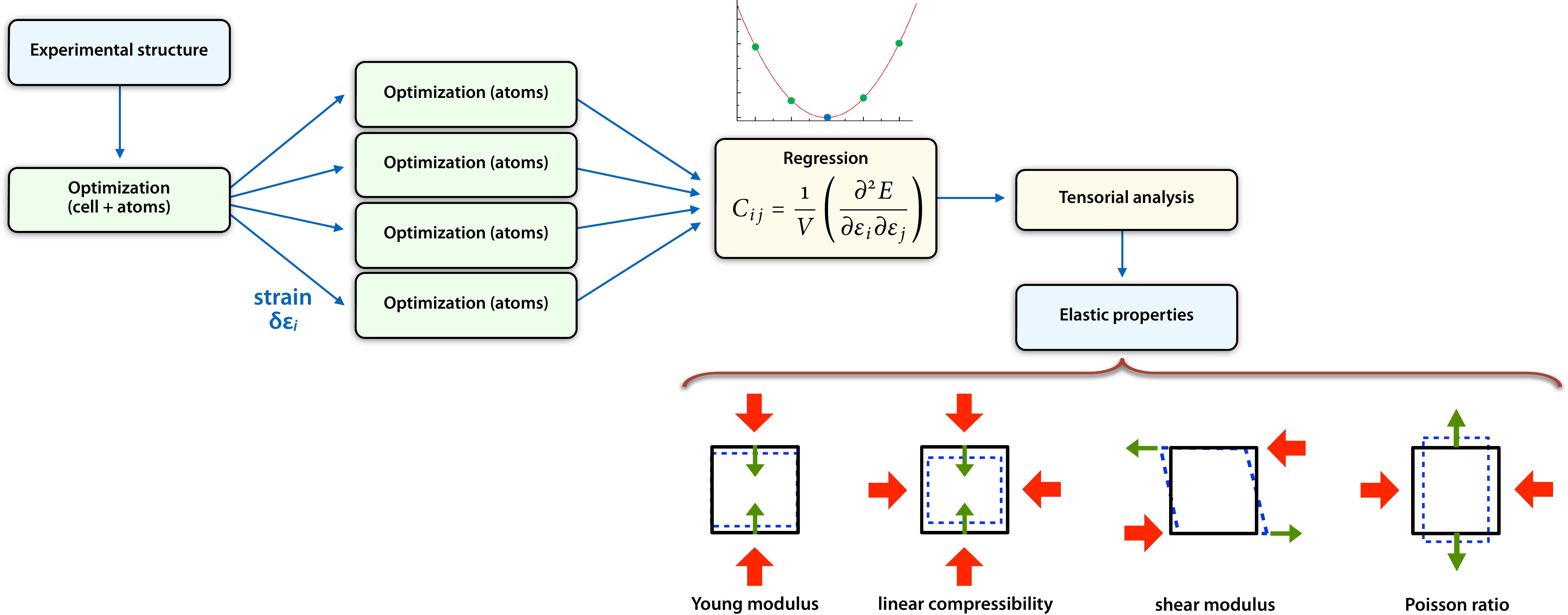}
\caption{\label{fig:elastic}Determination of the four-rank tensor $\mathsf{C}$ of second-order elastic constants, by quantum chemistry calculations, and its analysis to obtain physical properties such as Young's modulus ($E$), shear modulus ($G$), linear compressibility ($\beta$), and Poisson's ratio ($\nu$).}
\end{figure*}

Characterization of elastic constants of MOFs can also be performed on crystal structures without cubic symmetry, with the same techniques. Although the manual generation of strained structures by hand makes it more tedious in lower-symmetry crystal classes, some quantum chemistry software now handle it directly (including CRYSTAL14\cite{CRYSTAL, Perger2009}), and scripts are available for others. These allow to calculate the full \textbf{four-rank tensor $\mathsf{C}$ of second-order elastic constants}, which relates strain $\epsilon$ to stress $\sigma$ in the tensorial Hooke's law: \begin{equation} \sigma_{ij} = \sum_{kl} \mathsf{C}_{ijkl} \epsilon_{kl} \end{equation} The number of independent nonzero elements of this stiffness tensor, depends on the crystal class. It determines the entire elastic behavior of the material, and by tensorial analysis can be used to calculate physical properties of interest (see Figure~\ref{fig:elastic}), including: \begin{itemize} \item the directional Young's modulus $E(\mathbf u)$, also known as the tensile modulus, quantifies the deformation of the material in direction $\mathbf u$, when it is compressed in that same direction; \item the linear compressibility $\beta(\mathbf{u})$, which characterizes the compression along axis $\mathbf{u}$ when the crystal undergoes an isotropic compression; \item the shear modulus (or modulus of rigidity) $G(\mathbf u, \mathbf v)$, which quantifies the material's response to shearing strains along $\mathbf{u}$, in the plane normal to $\mathbf{v}$; \item the Poisson's ratio $\nu(\mathbf{u},\mathbf{v})$ which characterizes the transverse strain (in the $\mathbf{v}$ direction) under uniaxial stress (in the $\mathbf{u}$ direction). \end{itemize} Programs are available for the calculation of these properties from the stiffness tensor, such as the ElAM\cite{Elam} code or the online ELATE web app.\cite{ELATE}

\begin{figure}[t]\centering
\includegraphics[width=\linewidth]{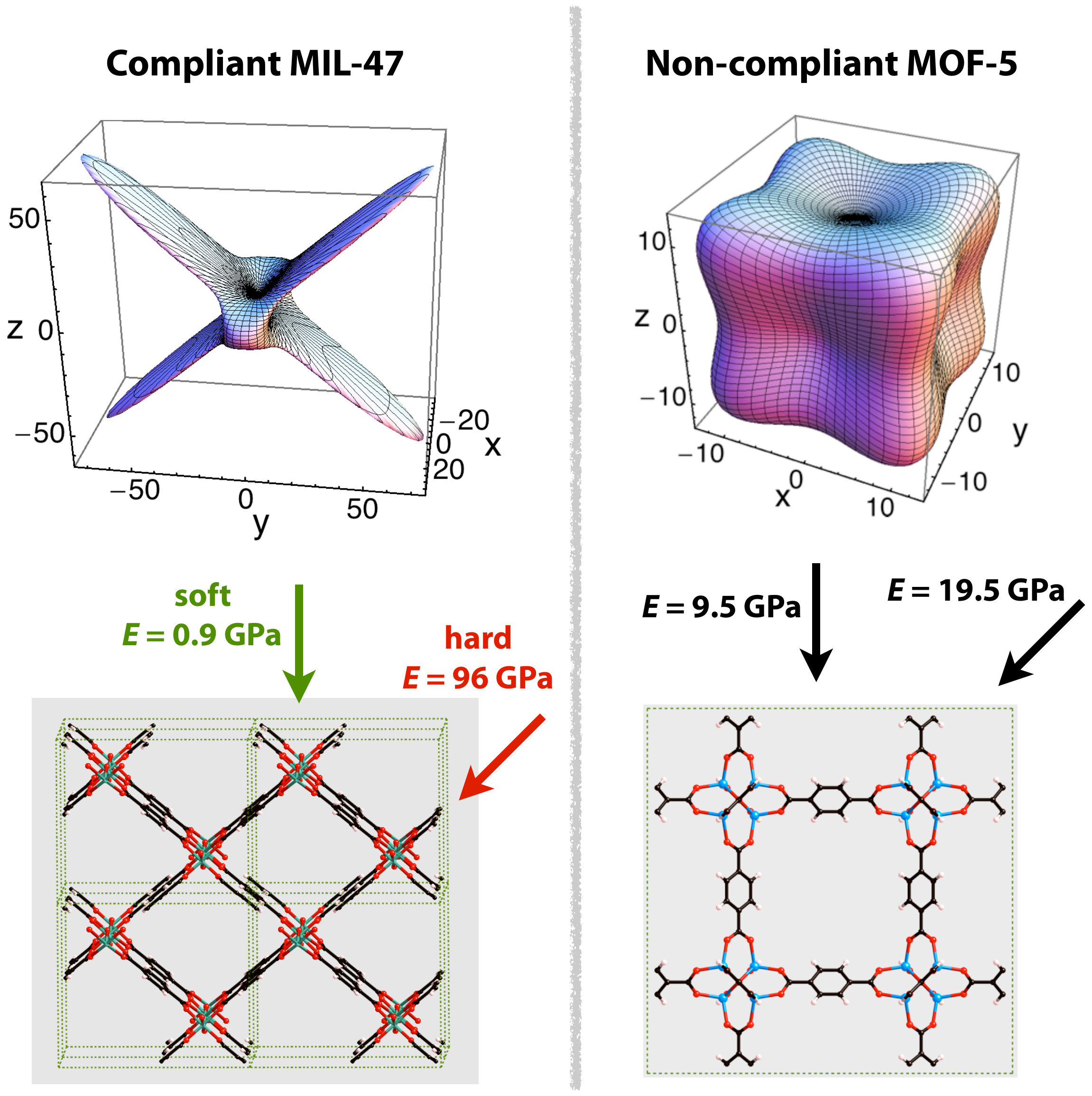}
\caption{\label{fig:elastic_SPC}Anisotropy of the elastic moduli as a signature of \emph{soft porous crystals}: 3D representation of the directional Young's modulus of ``breathing'' metal--organic framework MIL-47, compared to MOF-5. On the bottom panel are indicated the stiffest and softest directions of deformation (minimal and maximal Young's modulus). Adapted with permission from ref.~\citep{Coudert2013_perspective}. Copyright 2013 American Chemical Society.}
\end{figure}

The detailed analysis of the elastic properties of metal--organic frameworks has, in the last few years, been applied to different properties. In addition to the low elastic moduli of MOFs in general, Ortiz et al.\cite{Ortiz_PRL2012, Ortiz_JCP} demonstrated that the existence of high anisotropy in elastic properties, coupled with directions of very small Young's and shear moduli (sub-GPa), is a \textbf{signature of the flexibility of \emph{soft porous crystals}},\cite{Kitagawa_SPC} determining their ability to undergo large structural deformations under stimulation (see Figure~\ref{fig:elastic_SPC}).\cite{Coudert2015} This has allowed the prediction of flexibility for new structures, such as NOTT-300 and CAU-13\cite{Ortiz2014_ChemComm} (the latter has since been confirmed experimentally\cite{Niekiel2014}).

Among the mechanical properties of MOFs that have attracted some interest, \textbf{Negative Linear Compressibility} (NLC)\cite{Cairns2015} has drawn significant attention. This property of materials expanding in one or two directions under hydrostatic compression is considered rather exotic in inorganic solids,\cite{Miller2015} but has been observed in a relatively large number of framework materials\cite{Goodwin2008, Fortes2011, Cairns2013, Gatt2013} and MOFs.\cite{Li2012_JACS, SerraCrespo2015, Cai2014} On the computational side, it can be studied in two ways. In the linear elasticity regime, it can be studied through the computation of the elastic stiffness tensor as described above, as was done for the prediction and subsequent experimental confirmation of negative linear compressibility in the MIL-53  family.\cite{SerraCrespo2015} It can also be performed by \emph{in silico} compression experiments, studying the influence of finite increments of pressure on a MOF structure, either through enthalpy minimization calculations under pressure,\cite{Li2012_JACS} or through constant-pressure constant-temperature $(N, \sigma, T)$ molecular dynamics studies.\cite{Ortiz2014_ChemComm}

\begin{figure}[t]\centering
\includegraphics[width=\linewidth]{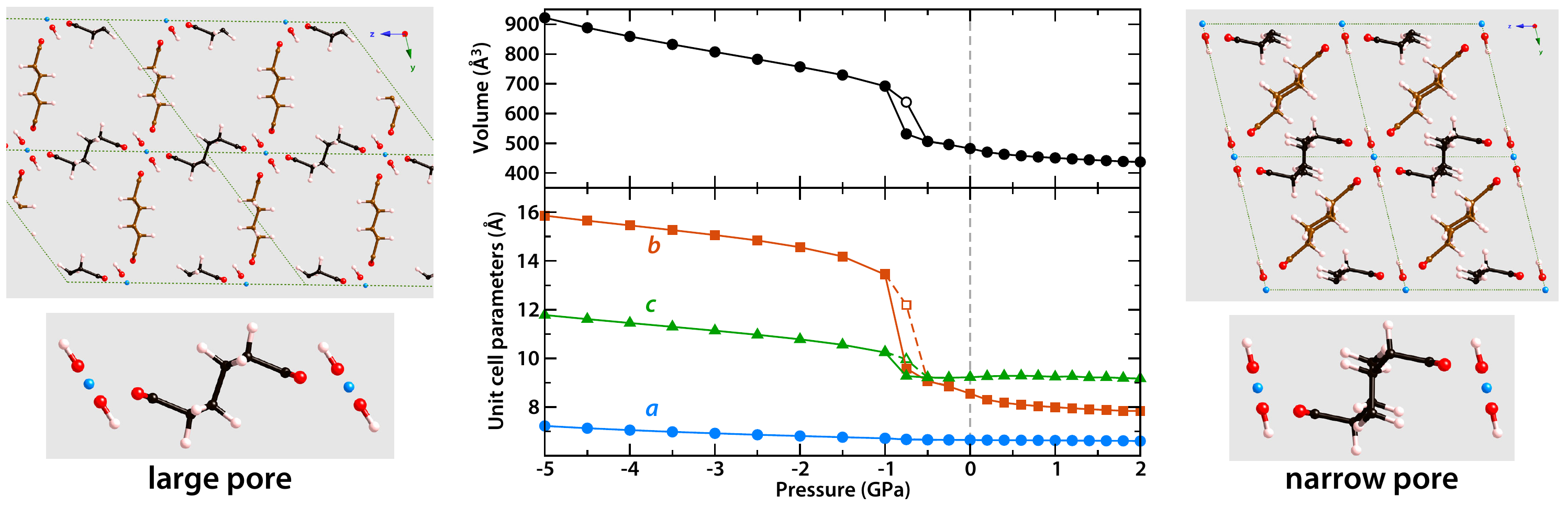}\\\rule{0.7\linewidth}{1pt}\\[3mm]
\includegraphics[width=\linewidth]{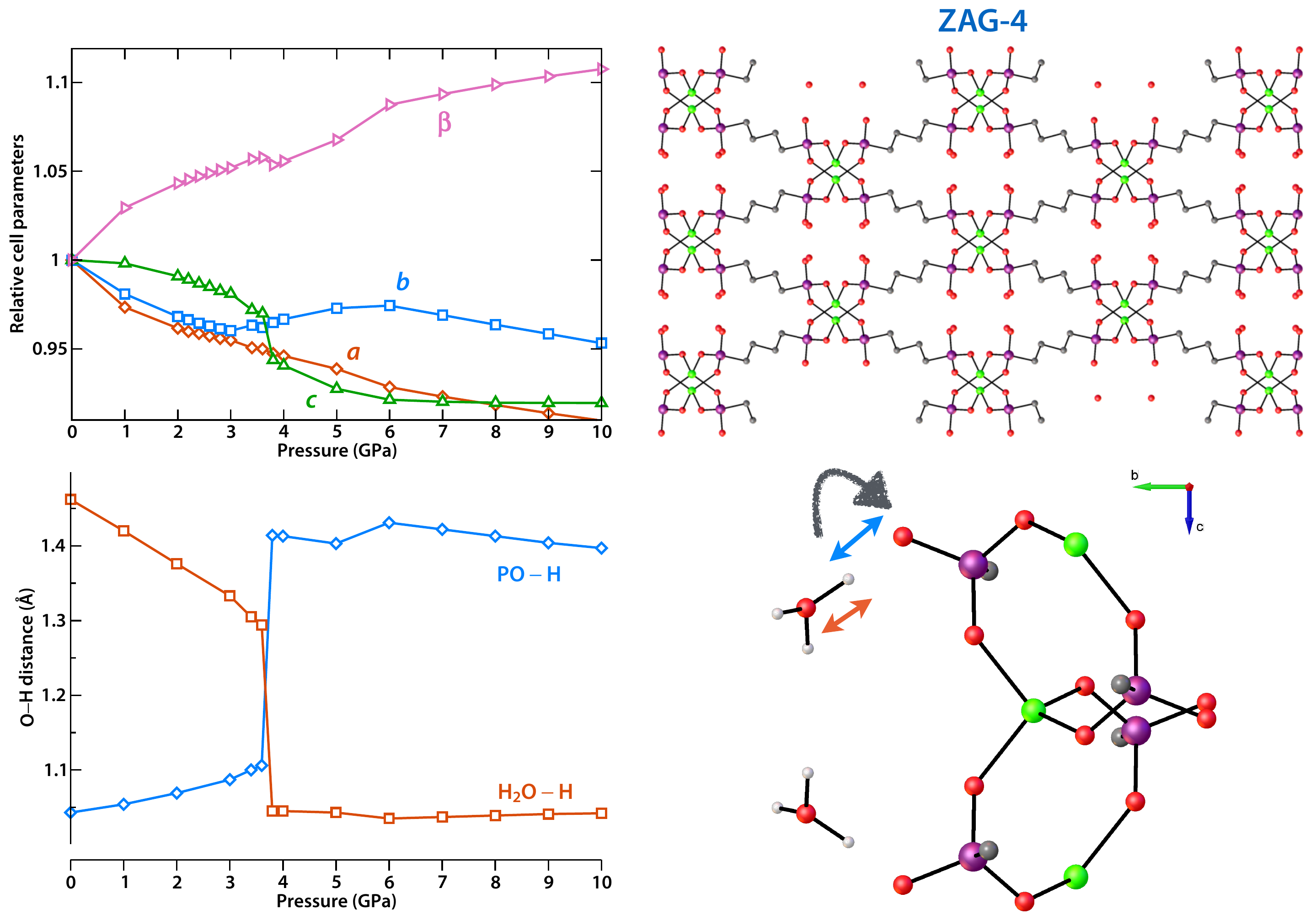}
\caption{\label{fig:compression}Examples of \emph{in silico} compression experiments, by DFT-based enthalpy minimization under pressure. Top: narrow-to-large-pore structural transition in CAU-13 under tension. Bottom: pressure-induced proton jump in ZAG-4. Adapted with permission from refs.~\citep{Ortiz2014_ChemComm} (from The Royal Society of Chemistry) and \citep{Ortiz2014_JACS} (Copyright 2014 American Chemical Society.)}
\end{figure}

Because computational compression experiments allow the determination of structure (unit cell parameters and atomic positions) as a function of applied mechanical pressure, it yields information outside of the elastic regime and can provide insight into the occurrence of \textbf{pressure-induced structural transitions}. Two rather typical examples of these \emph{in silico} compressions are shown in Figure~\ref{fig:compression}, using DFT-based enthalpy minimization of the crystal structures at increasing (or decreasing) values of pressure. The first is that of CAU-13,\cite{Niekiel2013} demonstrating linear elastic behavior at positive pressure and the existence of a narrow-pore-to-large-pore structural transition under tension, i.e. for hydrostatic pressures around $P \simeq -500$~MPa. Although experimentally applying hydrostatic tension is not an option, it does correspond to the outward stress induced by adsorption of bulky molecules (such as xylenes) in small nanopores. The second example is that of ZAG-4, a Zinc Alkyl Gate material showing nonmonotonic behavior under compression, as seen from high-pressure single-crystal X-ray crystallography.\cite{Gagnon2013} Quantum chemical calculations of the compression process showed that the nonlinear behavior is associated with a structural transition, namely a reversible pressure-induced proton transfer between an included water molecule and the linker's phosphonate group.\cite{Ortiz2014_JACS}

Finally, it should be noted that in the (relatively rare) case of materials for which a good flexible force field is available, such studies can also be performed using force field-based molecular dynamics simulations. This approach has been well demonstrated in the case of the crystal-to-crystal ``breathing'' transition in materials of the MIL-53 family (including MIL-47), as is discussed fully in Section~\ref{sec:flexible}. The same approach was used to study the pressure-induced amorphization in ZIF-8 and ZIF-4.\cite{Ortiz2013} There, fully-anisotropic constant-pressure molecular dynamics simulations were used to calculate the evolution of elastic constants of the materials at room temperature, as a function of pressure.\footnote{The elastic constants $C_{ij}$ can be obtained from the fluctuations of the unit cell vectors in constant-stress $(N, \sigma, T)$ molecular dynamics simulations with anisotropic variations of the unit cell, using the strain-fluctuation formula:\cite{Parrinello1982} \[ \left(\frac{kT}{V}\right) C_{ij}^{-1} = \left<\epsilon_i\epsilon_j\right> - \left<\epsilon_i\right>\left<\epsilon_j\right> \]} This was used to show that the pressure-induced amorphization of both ZIF-8 and ZIF-4 is due to a shear mode softening under pressure,\cite{Ortiz2013} leading to mechanical instability at sub-GPa pressures when the Born stability conditions are no longer satisfied.\cite{Mouhat2014} This approach was later extended to a larger database of ZIF structures, showing the very limited mechanical stability of a majority of both hypothetical and experimental structures upon solvent or guest evacuation.\cite{BouesselduBourg2014}

\subsection{Thermal properties}

Because of its scalar nature and limited practical range, the diversity in MOF response to temperature is somewhat narrower than their responses to pressure, described above. Yet there has been a relatively large number of computational studies on the thermal properties of MOFs, and in particular their thermal expansion. There is, among the whole class of metal--organic frameworks, a prevalence of \textbf{negative thermal expansion (NTE)} and its occurrence in relatively large temperature ranges often including room conditions. Materials exhibiting NTE contract when heated, a rare property among dense inorganic materials and usually limited to certain types of structures.\cite{NTE} Negative thermal expansion, however, is quite common among molecular frameworks, and has been observed experimentally in many metal--organic frameworks, including HKUST-1,\cite{Wu2008} MOF-5,\cite{Zhou2008} other members of the IRMOF family,\cite{Dubbeldam2007} and many ZIFs.\cite{BouesselduBourg2014}

One possible way to study the thermal expansion of MOFs is to perform constant-pressure constant-temperature ($N, \sigma, T$) molecular dynamics simulations for various values of temperature. The evolution of the volume and unit cell parameters then allow a direct determination of both the volumetric and linear thermal expansion coefficients, $\alpha_V$ and $\alpha_\ell$ respectively: \begin{equation} \alpha_V=\frac{1}{V}\left(\frac{\partial V}{\partial T}\right)_{N, \sigma} \ \ ; \ \ \alpha_\ell=\frac{1}{\ell}\left(\frac{\partial \ell}{\partial T}\right)_{N, \sigma} \end{equation} where $\ell$ is any unit cell parameter. This was used to confirm the existence of NTE in several materials of the IRMOF family (using force field-based MD simulations), as well as to shed light onto its microscopic origin.\cite{Han2007, Dubbeldam2007} The propensity of IRMOFs for NTE was attributed to the presence of many soft (low frequency) transverse vibrational modes in the frameworks, and in particular the vibration modes of the linkers transverse to the metal--metal axes, resulting in a shorter average effective length of the linkers, even though all the bond lengths in the structure increase with temperature. In a later study, Peterson et al. combined neutron scattering with \emph{ab initio} molecular dynamics of the metal clusters (copper paddle wheels) of HKUST-1 to elucidate the origin of its NTE, involving both concerted transverse vibrations as well as local molecular vibrations such as paddle wheel twisting deformation.\cite{Peterson2010}

\begin{figure*}[t]\centering
\includegraphics[width=0.7\linewidth]{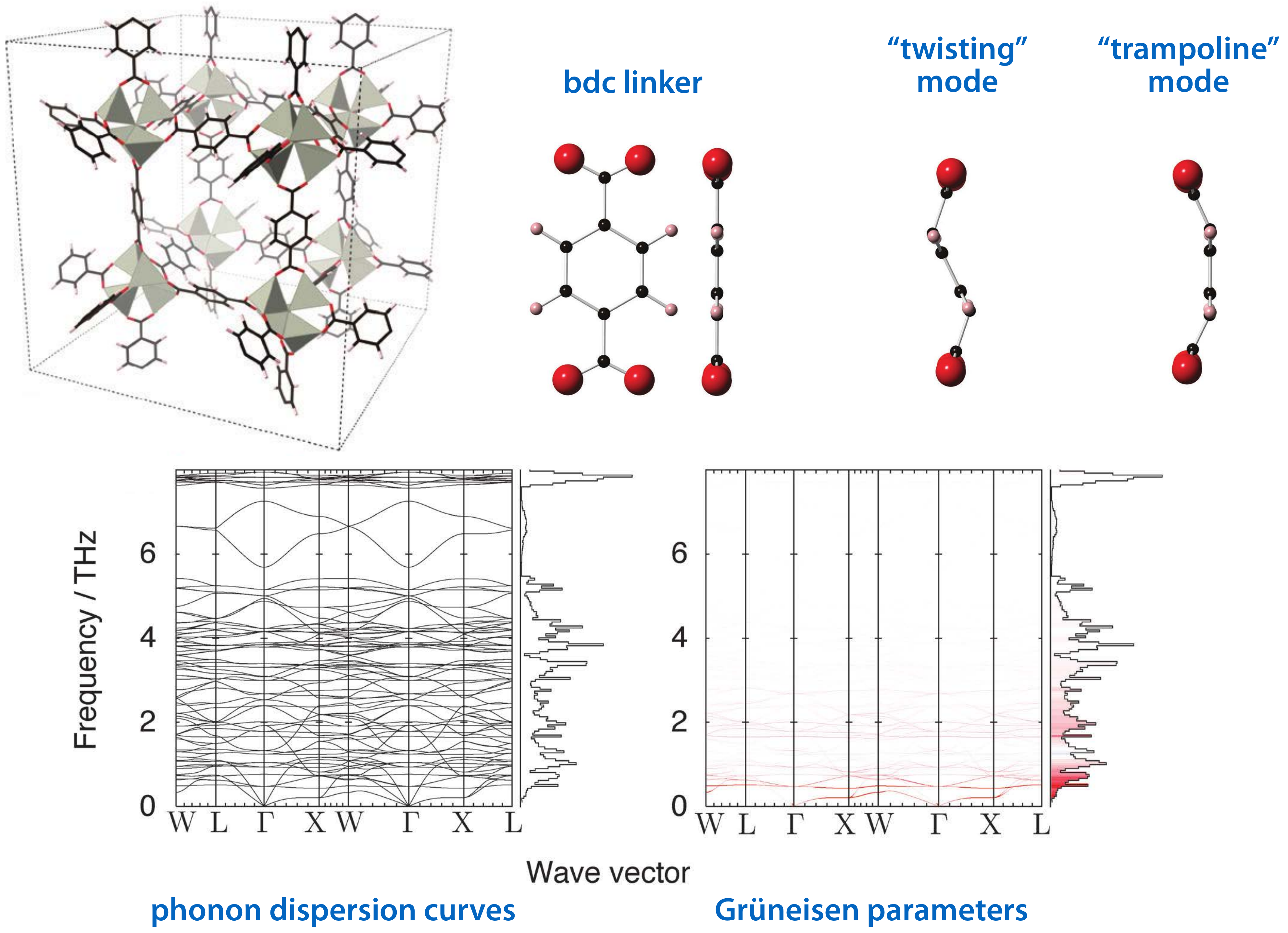}
\caption{\label{fig:NTE_phonons}Top: MOF-5 unit cell and main NTE-contributing vibration modes of the linkers. Bottom: phonon dispersion curves and density of states (bottom left), and associated values of Gr\"{u}neisen parameters (bottom right; redder = more negative). Adapted from ref.~\citep{Rimmer2014} with permission from The Royal Society of Chemistry.}
\end{figure*}

Another approach to study the lattice dynamics of MOFs and their thermal expansion is to calculate the thermal expansion coefficients within the quasi-harmonic approximation. This requires the determination of the phonon modes and frequencies of the structure, and their dependence on unit cell parameters: this is typically done by \emph{ab initio} lattice dynamics calculations of the phonon frequencies at various points throughout the Brillouin zone. For each phonon mode $i$ and $\mathbf{k}$ point, a mode-specific Gr\"{u}neisen parameter, $\gamma_{i,\mathbf{k}}$ can be calculated: \begin{equation} \gamma_{i,\mathbf{k}} = -\left(\frac{\partial\log \omega_{i,\mathbf{k}}}{\partial\log V}\right) \end{equation} The phonon modes with both negative Gr\"{u}neisen parameter and low frequency (and thus large amplitude) drive the negative thermal expansion behavior. This approach was used, for example, to study the NTE in MOF-5, first at the $\Gamma$-point only,\cite{Zhou2008} then later across the full Brillouin zone.\cite{Rimmer2014} This allowed to identify both optic and accoustic modes contributing to the macroscopic NTE of MOF-5 (as schematized in Figure~\ref{fig:NTE_phonons}). Unlike the direct molecular dynamics approach, the quasiharmonic approximation cannot account for high anharmonicity, but it does give more insight into the roots of the negative thermal expansion at the microscopic level.

Finally, another thermal property of interest in MOFs is their \textbf{heat capacity}, an important thermodynamic parameter to characterize the material itself and also optimized the adsorption process for practical applications. Porous solids with larger heat capacity can better adsorb the heat generated during adsorption, thus limiting the undesirable thermal effects (increased temperature leading to lower adsorption). However, very few papers directly address this issue, whether experimentally\cite{Mu2011} or computationally.\cite{Bristow2014}

\subsection{Optical and electronic properties}

\begin{figure*}[t]\centering
\includegraphics[height=42mm]{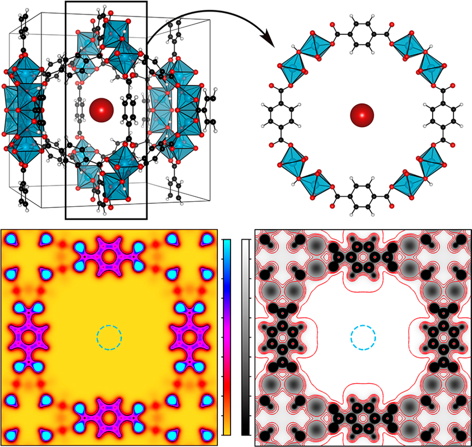}\hfill
\includegraphics[height=42mm]{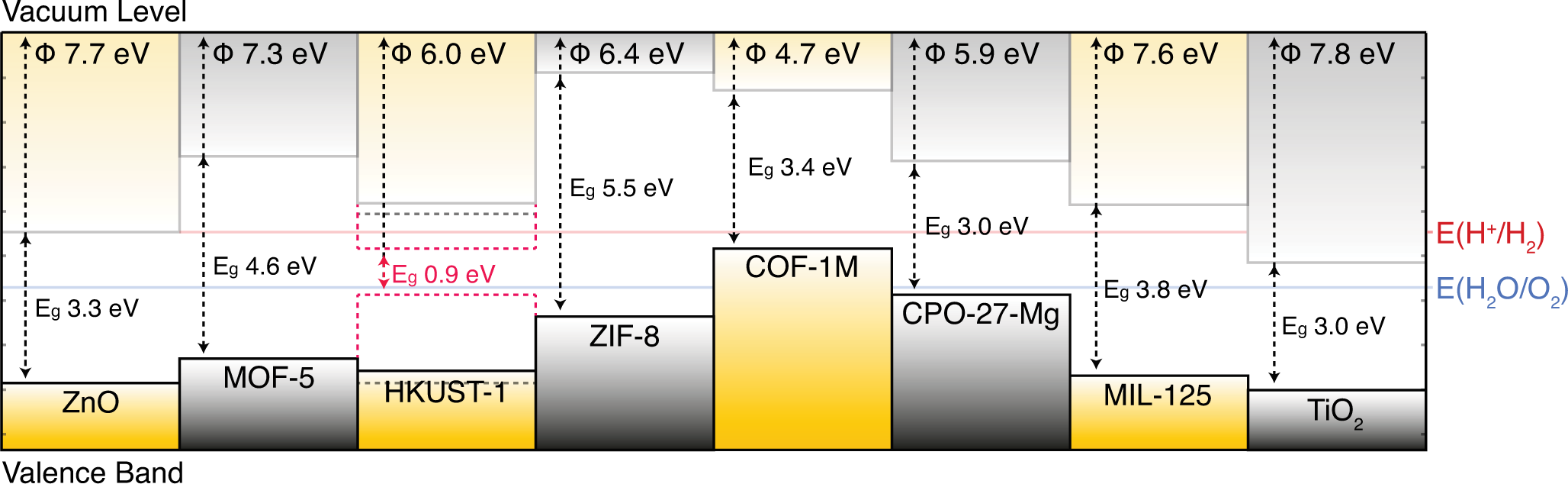}
\caption{\label{fig:Butler}Left: calculation of MIL-125's electrostatic reference potential for MOFs, depicting the reference point chosen at the pore center. Right: vertical ionization energy of six porous MOFs with respect to a common vacuum level. Reproduced with permission from ref.~\citep{Butler2014}. Copyright 2014 American Chemical Society.}
\end{figure*}

Optical properties of metal--organic frameworks, including UV-visible absorption and luminescence, have attracted a great deal of attention due to their potential for applications in chemical, biological, and radiation detection, medical imaging, and electro-optical devices.\cite{Allendorf2009, Cui2012} In addition, there is also great interest in the characterization and control of band gaps, for applications in electronic devices, solar energy harvesting and photocatalysis.\cite{Silva2010, Allendorf2011, Wang2011} These optical and electronic properties of MOFs, require modeling at the quantum chemical level, mostly in the Density Functional Theory (DFT) approach, in order to fully characterize the electronic structure of the material.

A lot of the early work on the modeling of electronic properties of MOFs focused on the characterization of their \textbf{band gaps},\cite{Mattesini2006} and the possibilities for tuning these by metal exchange\cite{FuentesCabrera2005, Yang2014} or ligand functionalization\cite{FlageLarsen2013} or substitution.\cite{Civalleri2006} Probably one of the most comprehensive examples of this kind of studies is that of Hendon et al,\cite{Hendon2015} who explored through a combination of synthetic and computational work the influence of linker functionalization on the band gap of MIL-125, a photochromic MOF based on TiO\e{2} and 1,4-benzenedicarboxylate.\cite{DanHardi2009} By studying materials built from linkers with various functional groups, the authors demonstrated that the diaminated linker bdc-(NH\e{2})\e{2} was the most efficient in lowering the band gap, from 3.6 to 1.3~eV. They also demonstrated that the introduction of just one aminated linker per unit cell was sufficient in lowering the band gap.

In addition to the value of band gap, more recent work has shifted focus to a broader range of electronic properties, including detailed analysis of electron density and electrostatic potential. One such example is the recently proposed method by Butler et al. to report \textbf{vertical ionization energy} of MOFs, with respect to a vacuum level set at the value of the electrostatic potential (for MOFs whose pores are large enough).\cite{Butler2014} Based on band gap and ionization potential values, the authors explain certain electrochemical, optical, and electrical properties of the materials studied. This method was also used for the characterization of the piezochromism of MOFs, i.e. their ability to change electronic and optical properties as a function of applied pressure.\cite{Butler2014b, Ling2015}

Finally, another area of interest is that of the luminescence of MOFs. The experimental literature on the topic (see for example the reviews in Refs.~\citenum{Allendorf2009}, \citenum{Cui2012} and \citenum{Hu2014}) largely outweighs the computational studies, and this is an area still very much in development. Several studies have attempted to determined the electronic nature of the absorption and emission transitions in MOFs, and their dependence on linker functionalization, nature of the metal center, and linker-linker interactions and stacking. The accurate determination of the luminescence properties require the use of computational intensive time-dependent density functional theory (TDDFT)\cite{tddft2006, Adamo2013} with \emph{ad hoc} exchange--correlation functionals and large basis sets in order to describe the excited states and optical (absorption and emission) spectra of MOFs. This has allowed to determine the nature of the emission transition in a few luminescent MOFs,\cite{Zhang2012_JMaterChem, Ji2013} including archetypical MOF-5.\cite{Ji2012}

\section{Adsorption\label{sec:adsorption}}

As nanoporous materials, adsorption of molecular fluids inside MOFs is one of their earliest and most studied properties. Due to their well-defined crystalline structure, high pore volume, large surface area and tailorable pore sizes, MOFs show great promise in becoming the next generation of nanoporous adsorbents. They are thus natural candidates to supplement or replace zeolites in adsorption-based industrial applications, with a specific focus on energy-related and biomedical applications. This includes drug delivery, gas adsorption and capture, gas storage and delivery, separation in gas and liquid phase, purification, and sensing.\cite{Li2009, Li2011, Li2012} In particular, the separation, capture, and sequestration of carbon dioxide from industrial and automotive emissions has recently attracted intense research interest in the effort to curb emissions of this greenhouse gas linked to human-induced global warming.\cite{Liu2012, Sumida2012}

Adsorption properties of new porous materials are routinely reported along with the synthesis, structure and other physical and chemical characterization. In addition to the standard nitrogen or argon adsorption at cryogenic temperatures, part of the BET measurement of surface area,\cite{Brunauer1938, IUPAC_porous} adsorption and desorption isotherms of gases of strategic interest such as CO\e{2}, CH\e{4}, CO, N\e{2}, and O\e{2}, is often performed to assess novel materials. Somewhat more specialized topic, which are nonetheless of practical importance, such as adsorption of hydrogen,\cite{Han2009, Murray2009, Suh2012, Yan2014} polar molecules (such as water\cite{Canivet2014, Burtch2014} and alcohols\cite{Wu2012}), and larger molecules (including hydrocarbons\cite{Lamia2009}) have all been extensively studied in some of the materials.

Given the importance of this field, there has been significant computational work addressing both the fundamental understanding of adsorption in metal--organic frameworks and the practical applications. Molecular simulation of adsorption allows to shed light into the microscopic root of the behaviors observed experimentally. It can also be a powerful tool for the quantitative \emph{prediction}  of adsorption and coadsorption in MOFs, as well as the optimization of materials for adsorption-based processes. Reviews on this topic are refs. \citep{Duren2009} and \citep{Getman2012}. Here, we summarize the state of the art in molecular simulation of adsorption in metal--organic frameworks, starting with the classical Grand Canonical Monte Carlo methods for adsorption and coadsorption of fluids, before listing advanced techniques developed in the last few years for the study of specific aspects of MOF adsorption.

\begin{figure*}[t]\centering
\includegraphics[width=0.7\textwidth]{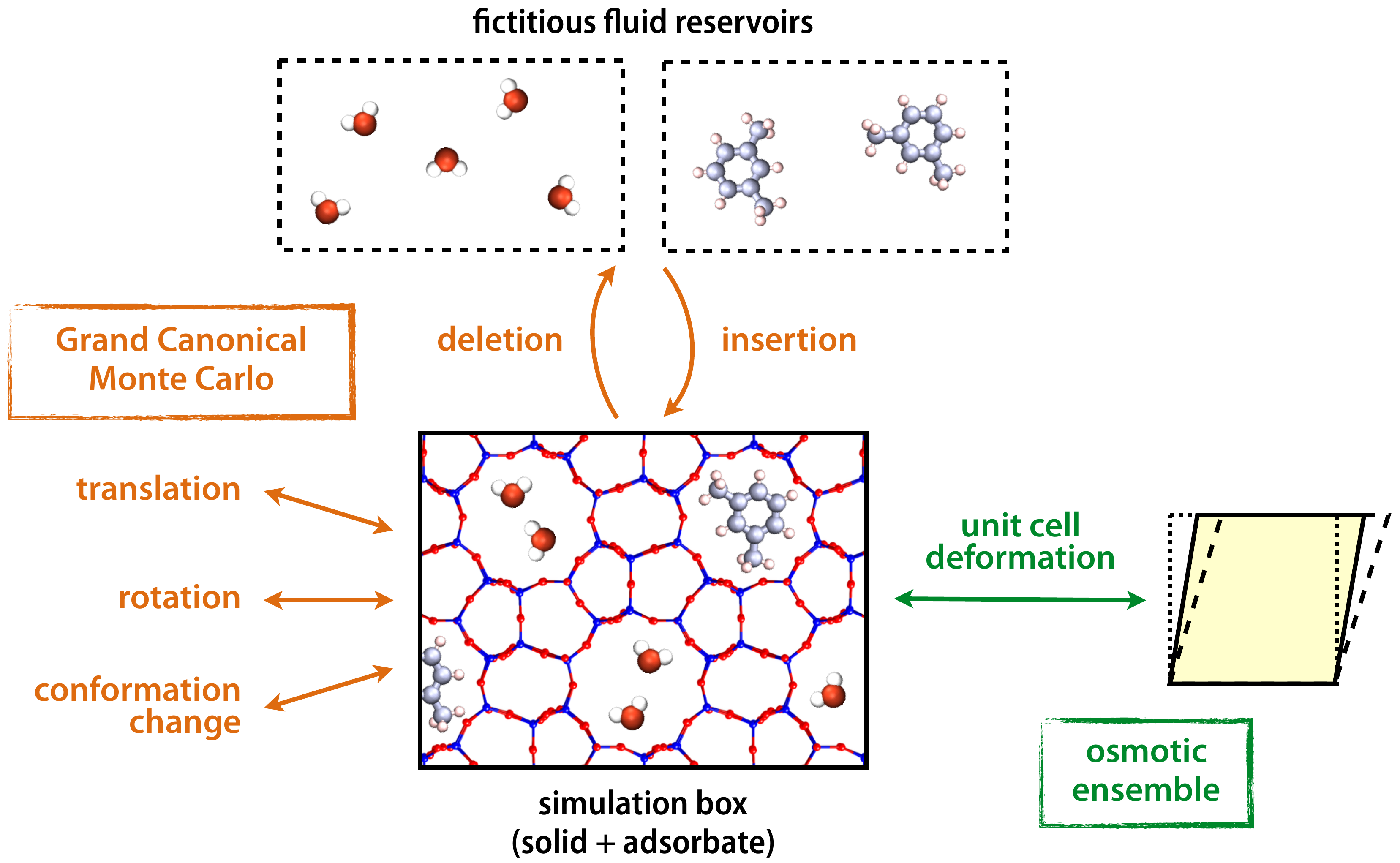}
\caption{\label{fig:MC}Schematic diagram of the Monte Carlo simulation of adsorption in a porous host, in the Grand Canonical thermodynamic ensemble (in orange) and in the osmotic ensemble (allowing deformation of the material, in green).}
\end{figure*}

\subsection{Grand Canonical Monte Carlo\label{sec:gcmc}}

The standard molecular simulation technique to study the thermodynamics of adsorption in rigid nanoporous materials is to use Monte Carlo simulations in the Grand Canonical ensemble, or Grand Canonical Monte Carlo (GCMC).\cite{Fuchs2001, Smit2008, Ungerer2005} It has been extensively used and validated in a large variety of nanoporous materials, including zeolites, nanoporous carbons, mesoporous materials, etc. This approach, schematized in Figure~\ref{fig:MC}, models the adsorption of molecular fluids or mixtures inside a rigid porous matrix (system of fixed volume $V$), for given values of the temperature ($T$) and chemical potential of the fluid ($\mu$).

In Monte Carlo simulations, series of configurations of the system under study are generated in a stochastic process, by random moves accepted according to each configuration's Boltzmann probability. These molecular Monte Carlo moves typically include translation and rotation of molecules, as well as intramolecular displacements, i.e. changes in a given molecule's conformation.  Grand Canonical Monte Carlo simulations go beyond this by simulating the exchange of molecules between the interior of a material's pore space and an external reservoir of bulk fluid, with additional moves consisting of insertion of molecules into the pore volume, as well as deletion (moving them back to the reservoir). This allows the direct simulation of an open thermodynamic ensemble, with thermodynamic equilibrium between two phases (bulk and adsorbate) without an explicit interface.

The conditions of a Grand Canonical Monte Carlo simulation are close to the thermodynamic conditions during experimental adsorption measurements. Each point of an adsorption isotherm in GCMC is the result of a simulation at fixed $(\mu, V, T)$, and the full isotherm $N\e{ads}(\mu)$ is obtained by running simulations at different values of $\mu$. This is close to experimental isotherms, which are typically measured as $N\e{excess}(P)$, where $P$ is the pressure of the external fluid. Two differences between simulated and experimental adsorption isotherms need to be taken into account for quantitative comparison. The first is to relate the chemical potential $\mu$ (used in GCMC) to the pressure $P$ (measured experimentally) of the bulk fluid. This can be achieved through an equation of state, as $(\partial\mu/\partial P)_T = V\e{m}(P,T)$, the molar volume of the fluid. This equation of state for the fluid needs to be obtained from prior Monte Carlo simulations, or alternatively obtained from experimental data. In the specific case of low-pressure adsorption, the ideal gas law might be used, then equating pressure and fugacity: $\mu = \mu^0 + RT \ln(P/P^0)$. The second important difference between isotherms calculated through GCMC and measured experimentally is that between absolute and excess adsorbed properties, respectively. This can be significant in high-pressure experiment, and needs to be accounted for before any comparison between theoretical and experimental data.\cite{Myers2002}

Finally, this short introduction to Grand Canonical Monte Carlo can be concluded with the simple question: what can one compute with GCMC simulations? The output of GCMC includes both equilibrium thermodynamic quantities, as well as a representative set of configurations of the system in the given conditions. The macroscopic quantities include the absolute adsorption uptake ($N\e{ads}$), i.e. the average number of adsorbed molecules in the porous system, which is used in plotting adsorption isotherms. It also includes energetic quantities, such as the isosteric heat of adsorption ($q\e{st}$). These quantities can be directly compared to experimental data for validation, or used predictively as input for thermodynamic models of industrial adsorption-based processes, such as fixed-bed adsorption columns. In addition, Monte Carlo simulations also result in the generation of a sample of representative configurations of the system, from which structural information can be obtained. In adsorption, in particular, one can thus plot density distributions for adsorbed molecules, yielding microscopic insight into the adsorption mechanism.

\subsection{Classical interaction potentials}

Monte Carlo simulations, as described above, rely on the evaluation of the energy of each generated configuration of the system in order to evaluate its Boltzmann probability and accept or reject it. In short, the accuracy of the averages computed from the GCMC are directly dictated by the accuracy of the description used for the interactions of the molecules in the system. Because such energy calculations of single configurations will be made on the order of millions of times per GCMC simulation, quantum chemical calculation of the energy of each configuration is out of the picture. Thus, the interactions between the fluid molecules and MOF need to be described using classical models: interaction potentials, also called \emph{force fields}, are analytical functions of interatomic distances. In this approximation, we typically break down the host--guest and guest--guest interactions into terms with analytical expressions and simple physical meaning. At the intermolecular level, these terms include Coulombic interactions, long-range dispersion, short-range interatomic repulsion, polarizability, etc. A wide variety of functional forms are available to describe these intermolecular interactions, including the common Lennard-Jones and Buckingham potentials, as well as the Morse potential and the Feynman--Hibbs quantum effective potential, necessary for including quantum effects in adsorption of light gases such as hydrogen. Each ``force center'', on which these intermolecular potentials act, need not necessarily be an atom. Approaches such as United Atom (UA),\cite{Martin1998} and its refinement Anistropic United Atom (AUA),\cite{Ungerer2000} have shown that it is possible of grouping a functional group (CH, CH\e{2}, CH\e{3}, \ldots) into a single force center. In addition to the intermolecular terms, intramolecular terms may need to be considered for all but the smallest, rigid molecules. Intramolecular terms typically include stretching, bending and torsion potentials.

The choice of force fields is one between accuracy and transferability. On the one hand, designing \emph{ad hoc} force fields for every material and guest molecule studied is a very time consuming trial-and-error process, which some consider close enough to a black magic, but it can provide very accurate description of the interactions in the system. On the other hand, the so-called transferable (or universal) force fields, which describe the same types of atoms in many different materials with identical parameters, are much simpler to use and adapt to new MOF structures but their simplicity comes at the cost of a more approximate description of the potential energy surface.

When it comes to transferable force fields, the most used are UFF (Universal Force Field)\cite{UFF} and DREIDING\cite{DREIDING}. Both force fields are generic enough to provide potential parameters for elements throughout the periodic table, covering both the organic linkers and metal atoms of MOFs.\cite{Duren_COPS} Some groups have nevertheless worked on extending these universal force fields to handle more transition metals, or to improve the description of certain types of atoms (like the oxygen atoms in metal oxide clusters).\cite{Addicoat2014} In addition, when modeling the adsorption of polar guest molecules, these force fields need to be supplemented with atomic partial charges to describe the Coulombic interactions. Classically, several methods are available for the calculation of partial charges in molecular systems based on high-level quantum chemical calculations, grouped broadly in two families: population analysis or charge partitioning methods (Mulliken,\cite{Mulliken1955} Hirshfeld,\cite{Hirshfeld1977} Bader\cite{Bader1990}), and electrostatic potential fitting (such as CHelpG\cite{Breneman1990} and RESP\cite{Cieplak1995}). Electrostatic potential fitting methods are inherently more suited for the purpose of obtaining partial atomic charges for interatomic Coulombic potentials, but were long limited to nonperiodic systems. The past five years have seen a lot of developments of the state of the art in this area, with a new generation of electrostatic potential-based methods for periodic solids such as REPEAT (Repeating Electrostatic Potential Extracted ATomic)\cite{Campana2009} and DDEC (Density-Derived Electrostatic and Chemical charges),\cite{Manz2010} as well as their later refinements.\cite{Manz2012, Gabrieli2015}

Other approaches to the problem of atomic partial charges determination have been proposed. If one wants to perform rapid calculations at the detriment of quality of the results, classical approximations can be used in the form of charge equilibration methods.\cite{Wilmer2012, Kadantsev2013} These are particularly useful for high-throughput characterization of large numbers of materials. At the other hand of the spectrum, it is also worth noting that the determination of atomic charges can also be bypassed entirely, using the full electrostatic potential map in the material's unit cell, as determined from DFT calculations.\cite{Watanabe2011} This latter approach can become, however, computational quite expensive for materials with large unit cells.

While transferable force fields can lead to reasonable agreement in the case of adsorption of small nonpolar molecules in some MOFs, it should be noted again that they are a rather crude ``first order'' approximation of the interatomic interactions. Thus, there is a real need for a systematic methodology for developing reliable force fields for novel or hypothetical MOF structures. A large number of such methods have been proposed in the past few years, using computationally-demanding high-level quantum chemistry calculations as a basis for determining force field parameters. The case of partial charges has been discussed already above, but other terms can also be derived from quantum chemical calculations: short-range repulsion, dispersion, polarization interactions, etc. These terms are typically optimized by choosing a functional form beforehand, and optimizing all available force field parameters to match certain properties of the high-level calculations. Some methodologies focus on properties of the equilibrium structure, such as lattice parameters, atomic positions, forces (i.e. first derivatives of the energy with respect to atomic coordinates), vibration frequencies (i.e. second derivatives), bulk modulus (second derivative of the energy with respect to strain), etc. Other methodologies rely on fitting energy, and sometimes atomic forces, on a large sample of representative configurations of the system, themselves extracted from prior molecular simulations. This sampling can be performed with MD or GCMC simulation using universal force fields, or \emph{ab initio} molecular dynamics.

In conclusion, a lot of methodologies have been proposed for the development of MOF force fields from first principles, many with great success on limited number of materials. Some of the recent work in this area includes: MOF-FF,\cite{Bureekaew2013_MOFFF} BTW-FF,\cite{Bristow2014} QuickFF,\cite{Vanduyfhuys2012, Vanduyfhuys2015} among others.\cite{Han2012, Schmidt2015} However, none of them so far have demonstrated a decisive edge in universality or large-scale predictive power. Development of first-principles force fields remains for the most part an open challenge in the molecular simulation of MOF adsorption. More studies of systematic \emph{evaluation} of force field performance are needed,\cite{McDaniel2015} and other options for parameter determination (such as genetic algorithms\cite{Tafipolsky2009, Tafipolsky2010}) might be necessary to reach a truly universal methodology.

\subsection{Coadsorption and separation}

\begin{figure*}[t]\centering
\includegraphics[width=0.9\textwidth]{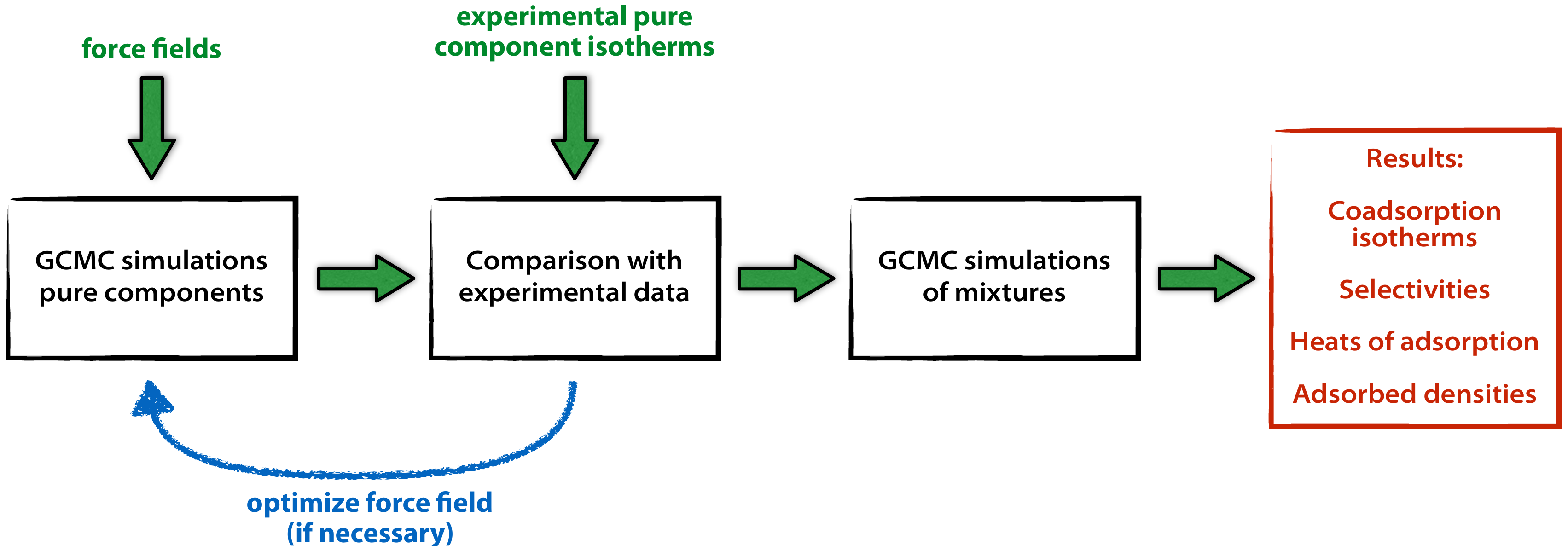}
\caption{\label{fig:coadsorption}Flowchart of a typical computational study of gas mixture coadsorption in metal--organic frameworks through GCMC simulations.}
\end{figure*}

\begin{figure*}[t]\centering
\includegraphics[width=\textwidth]{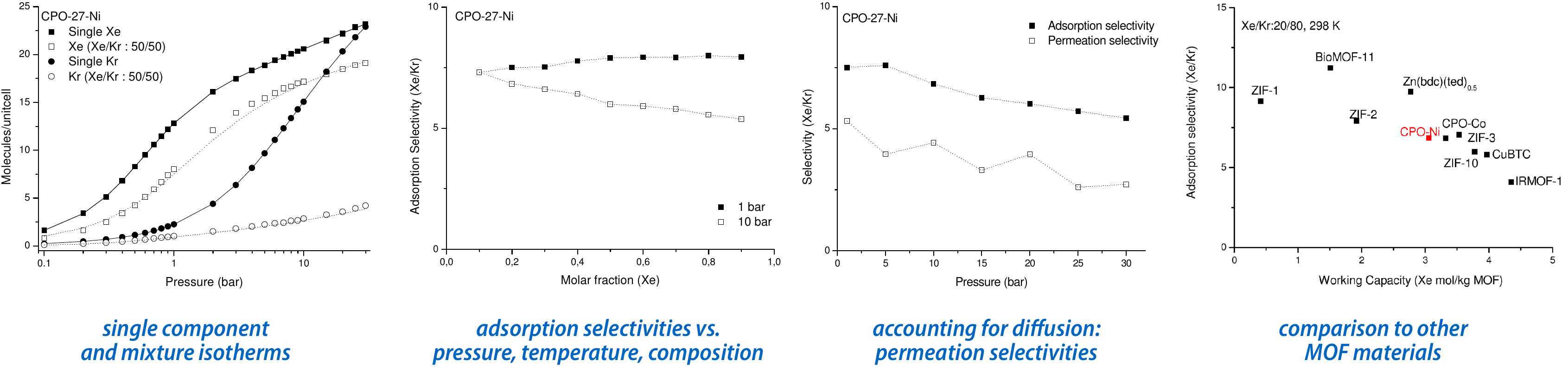}
\caption{\label{fig:coadsorption2}Results from a computational co-adsorption study of xenon/krypton mixtures in metal--organic framework CPO-27-Ni, by Gurdal et al.\cite{Gurdal2012} From left to right: single component and equimolar Xe/Kr mixture adsorption isotherms; adsorption selectivities at 1 and 10~bar, as a function of mixture composition; Xe/Kr adsorption and permeation selectivities as a function of pressure; performance of CPO-27-Ni for Xe/Kr separation compared to other MOFs. Adapted with permission from ref.~\citep{Gurdal2012}. Copyright 2012 American Chemical Society.}
\end{figure*}

The Grand Canonical Monte Carlo scheme, as described above, can model not only the adsorption of pure components, but also mixtures of molecular fluids, by specifying the chemical potential of each component in the mixture: $(\mu_1, \mu_2, \ldots, V, T)$. Like simulation of pure phases, Monte Carlo modeling of mixture coadsorption can be performed to provide microscopic insight into the driving forces for preferential adsorption or separation, for example by seeing how the adsorption of one component of the mixture enhances or suppresses the others, or by looking at guest--guest interactions.

But probably the greatest interest in the modeling of coadsorption through GCMC is in the computational prediction of fluid separation properties, which can be difficult and time-consuming to measure experimentally. This is particularly true of coadsorption because of the increased dimensionality of the problem, with a larger number of control parameters. Even for a binary mixture, coadsorption properties are a function of temperature, pressure and mixture composition, thus requiring tedious experimental work to map out in large ranges of each parameter. Even more so for ternary and more complex mixtures, with the number of experimental measurements for different compositions increasing exponentially with the number of components. There is thus great incentive in using molecular simulation of coadsorption to map out the separation properties of MOFs, finding optimal working conditions for applications, or providing thermodynamic information as input for process engineering and optimization. The process of a typically computational study of gas coadsorption through GCMC simulations of mixtures is schematized in Figure~\ref{fig:coadsorption}. Force fields (an input of the computation) are validated for each separate gas by comparison between experimental pure-component isotherms and GCMC simulations. If the agreement is not satisfactory, the MOF--adsorbate force fields need to be optimized: either adjusted empirically to the experimental data, or based on \emph{ab initio} reference data. Once the force fields have been validated for pure components, GCMC simulations of the mixture at various temperature, pressure and composition are performed. Analyzing these, the results obtained include coadsorption isotherms, selectivities, heats of (co)adsorption, plot of adsorbed densities for each component, etc. A rather typical example of this type of study is shown in Figure~\ref{fig:coadsorption2}, for the case of coadsorption of xenon and krypton in MOF CPO-27-Ni.\cite{Gurdal2012}

The other alternative to the modeling of coadsorption in MOFs, especially for gas mixtures, is the use of analytical mixture models such as the Ideal Adsorbed Solution Theory (IAST),\cite{IAST} or more complex variants such as Real Adsorbed Solution Theory (RAST)\cite{RAST, Hamon2012} or Vacancy Solution Theory\cite{VST, Wiersum2013}. While those have been shown to present reasonable predictions among mixtures of small molecules,\cite{Dickey2012, Yang2007, Bae2008} they can deviate from GCMC predictions and experiments in the cases of highly-competitive adsorption, tight packing of guest molecules (shape selectivity) or MOF structures with marked chemical heterogeneity.\cite{Keskin2009, Cessford2012} Thus analytical mixture theories, such as IAST, should only be used as a rough estimate of separation properties, before more time-consuming molecular simulations or experiments are performed.

\subsection{Beyond classical force fields: open metal sites and chemisorption\label{sec:abinitio}}

The molecular simulation of adsorption by Grand Canonical Monte Carlo relying on classical force fields is a very powerful tool, but it is limited by the accuracy of a given force field to describe host--guest interactions in the system. There are several cases where the classical approximations are not a realistic description of the intermolecular interactions, in particular in the case of chemisorption (i.e., formation of bond(s) between host and guest). Within the field of adsorption in metal--organic frameworks, the most common feature which classical force fields typically fail to describe is the short-range interaction of guest molecules with metal centers, i.e. the interaction with undercoordinated metal sites (or coordinatively unsaturated metal sites, \emph{cus}). This generic limitation has been evidenced on several pairs of adsorbents and guest molecules, including hydrogen,\cite{Fischer2010} methane,\cite{Chen2011} carbon dioxide,\cite{Yazaydin2009, Chen2012} propane and propylene,\cite{Fischer2012} etc.

In such cases, in order to accurately describe specific host--guest interactions (including guest interactions with coordinatively unsaturated metal sites), one needs to use \emph{ab initio} \textbf{quantum chemistry methods}, either in the Density Functional Theory (DFT) approach, using post-Hartree--Fock methods, or multi-reference methods. The latter two families of methods can be very accurate and require no \emph{ad hoc} parameters: they are thus very powerful to study novel systems for which classical force fields are not available, or where the adsorption mechanism is unknown, as well as very specific interactions. On this extensive literature, we refer the reader to the recent review of Odoh et al.\cite{Odoh2015} on quantum-chemical characterization of MOF properties, which includes an extensive section on the investigation of adsorption properties.

\begin{figure*}[t]\centering
\includegraphics[width=0.8\textwidth]{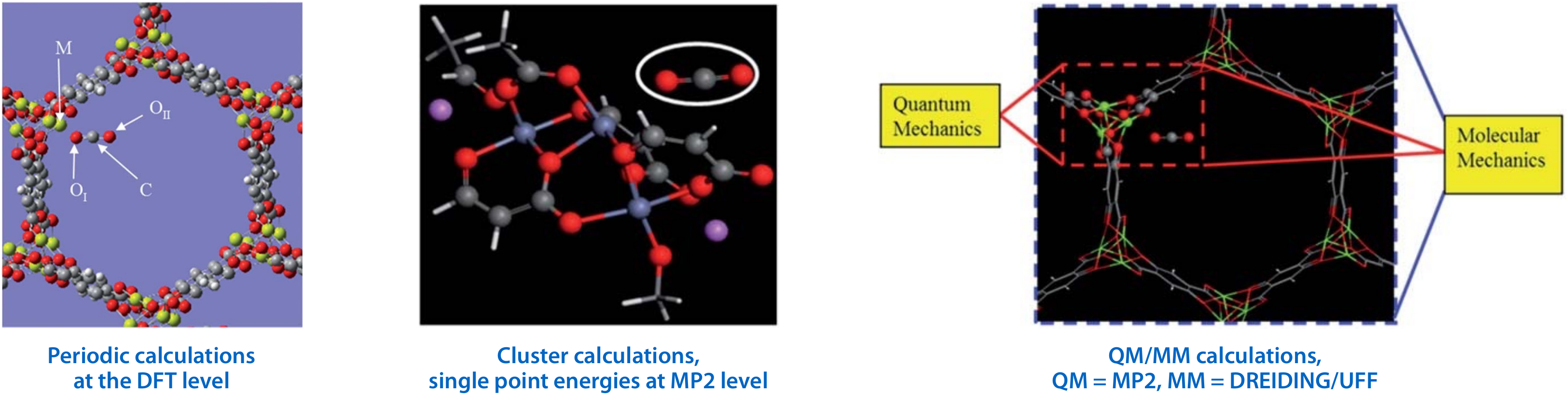}
\caption{\label{fig:Yu2013}Calculations of the CO\e{2} binding energies on open metal sites in CPO-27-M, with M~=~Mg, Mn, Fe, Co, Ni, Cu, or Zn. Adapted from ref.~\citep{Yu2013} with permission from The Royal Society of Chemistry.}
\end{figure*}

A rather representative example of this is the calculation of CO\e{2} adsorption in metal--organic frameworks of the CPO-27 family with different metals, using a combination of DFT and post-Hartree--Fock methods.\cite{Yu2013} Yu et al. have reported very accurate calculations of the CO\e{2} binding energies on open metal sites in CPO-27-M, with M~=~Mg, Mn, Fe, Co, Ni, Cu, or Zn. They first performed geometry optimizations of the periodic structures (atomic positions and lattice parameters) at the DFT level with generalized gradient approximation (GGA) exchange--correlation functional. Adsorption geometries were determined by geometry optimizations on representative clusters of the materials (see Figure~\ref{fig:Yu2013}), again at the DFT level. Accurate binding energies were then calculated using these geometries by single point energies at the MP2 level of theory. They were then further corrected with a QM/MM approach, where the QM energies were from the MP2-level calculations and the MM energies were calculated with the Dreiding and UFF force fields.

\begin{figure}[t]\centering
\includegraphics[width=0.8\linewidth]{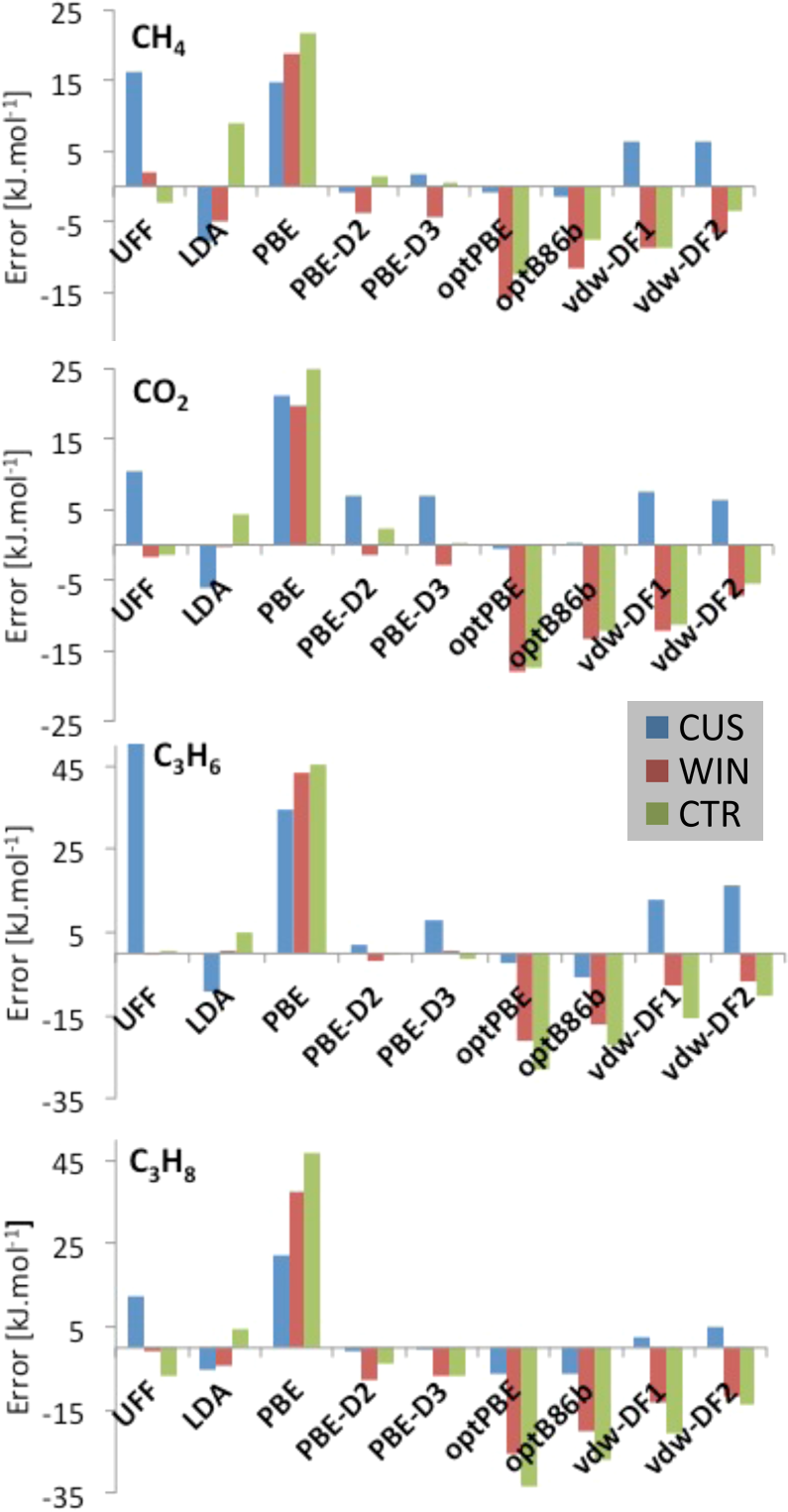}
\caption{\label{fig:Grajciar2015}Errors in the interaction energies calculated for methane, carbon dioxide, propane, and propene at CUS (open metal site), WIN (cage window), and CTR (cage center) sites of HKUST-1 with respect to the DFT/CC reference level of theory. Adapted with permission from ref.~\citep{Grajciar2015}. Copyright 2015 American Chemical Society.}
\end{figure}

In contrast with this high-accuracy methodology, studies of adsorption in MOFs with coordinatively unsaturated sites can also be performed entirely at the DFT level. These are computationally much cheaper and have been widely used in the literature. They need, however, to be carefully benchmarked against experimental data or highly accurate reference calculations, as the choice of DFT exchange--correlation function and dispersion correction scheme can heavily influence the properties calculated, such as  adsorption energies. There appears to be no ``one size fits all'' choice of methodology in this area, and optimal exchange--correlation functionals for each adsorbate/MOF pair need to be found by comparing the performance of various functionals on smaller cluster models using post-Hartree--Fock or multi-reference calculations.\cite{Rana2012, Ji2014, Grajciar2015} This is shown for example on Figure~\ref{fig:Grajciar2015}, a comprehensive plot of binding energies for several small molecules (methane, carbon dioxide, propane, and propene) in HKUST-1, evaluated with various DFT exchange--correlation functionals and compared to DFT/CC results.\cite{Grajciar2015}

Energy minimization calculations, as described above, only describe well-defined adsorption sites of a single molecule, as well as the corresponding geometries and low-coverage adsorption enthalpies (i.e., adsorption enthalpies in the limit of zero gas pressure). They do not account for entropy and guest--guest interactions, and thus cannot treat pore filling or packing effects of the molecules inside the pores. As a consequence, they cannot describe the adsorption at high uptake or where entropy plays a big role. Therefore, there has been a very large effort in the literature to use the insight and data gained from \emph{ab initio} calculations (adsorption sites, binding geometries, energies, forces, etc.) in order to \textbf{parameterize host--guest force fields}. This can take the form of reoptimizing existing force fields, adjusting some of the terms to match the quantum chemistry data: typically, strengthen the metal--guest potentials to more accurately describe the binding of molecules to the open metal sites of the framework. It can mean choosing an entirely altogether functional form for the metal--adsorbate interactions, or even a numerical description based on the energy profile as a function of adsorbate--metal center distance.

\begin{figure*}[t]\centering
\includegraphics[width=0.9\textwidth]{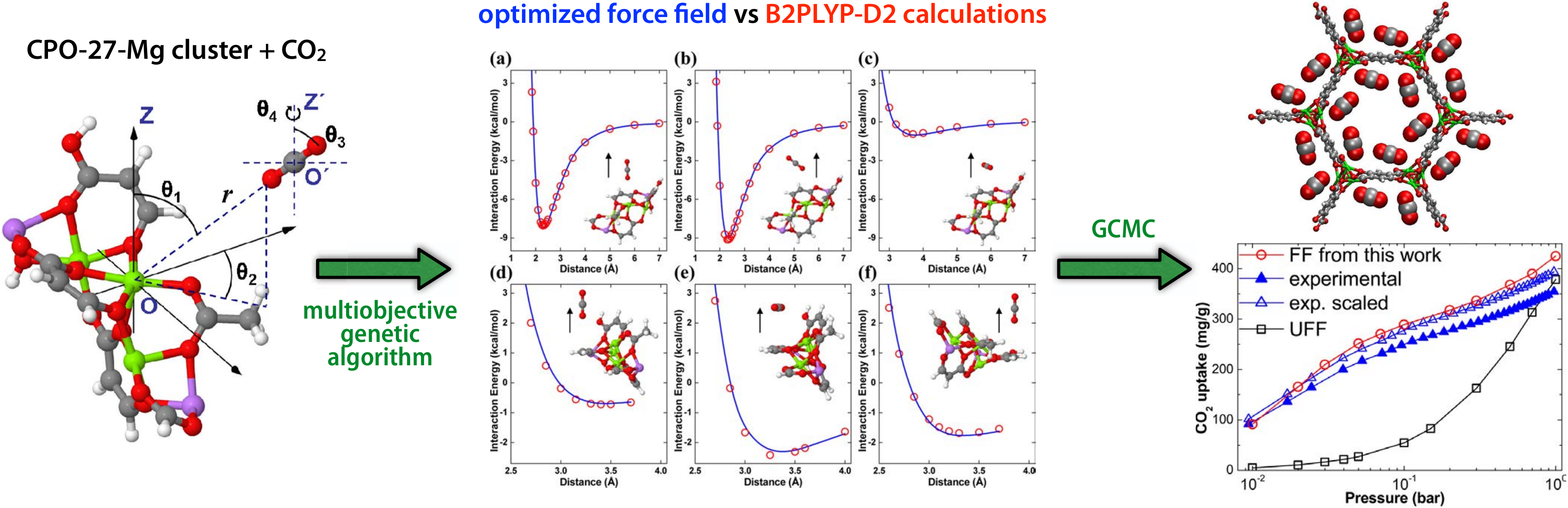}
\caption{\label{fig:Chen2012}\emph{Ab initio} parameterization of a force field describing the adsorption of CO\e{2} in CPO-27-Mg. Adapted with permission from ref.~\citep{Chen2012}. Copyright 2012 American Chemical Society.}
\end{figure*}

There are several examples of force field optimization for open metal site MOFs in the literature, including the cases of: hydrogen in HKUST-1;\cite{Fischer2010} water in copper-based MOFs;\cite{Zang2013} carbon dioxide and water in Mg-MOF-74\cite{Dzubak2012} and Zn-MOF-74;\cite{Lin2014_JCTC} CO\e{2} adsorption in Fe\e{2}(dobdc);\cite{Borycz2014} etc. We detail here by way of illustration the procedure followed by Chen et al.\cite{Chen2012} for the \emph{ab initio} parameterization of a force field describing the adsorption of CO\e{2} in CPO-27-Mg. Starting from a Buckhingham-type potential (the Carra--Konowalow potential), the authors calibrated a MMSV (Morse--Morse--spline--van der Waals) piecewise interaction potential for interactions with the MOF's coordinatively unsaturated
metal sites. The optimization of the MMSV potential was performed against reference \emph{ab initio} data obtained with a double-hybrid density functional with empirical dispersion correction, B2PLYP-D2.\cite{Grimme2006, GrimmeD2} The optimization itself was performed using a multiobjective genetic algorithm,\cite{Deb2009} and the good agreement between energy curves and \emph{ab initio} data is evident (see Figure~\ref{fig:Chen2012}). Chen et al. then used the optimized force field to perform GCMC simulations of adsorption isotherms, and showed vastly improved agreement compared to the standard force fields such as UFF.

Another solution that has been proposed is to precompute the MOF--guest interactions on a fine mesh of points within the pores of the MOF at the quantum chemical level. Using single-point quantum chemistry calculations, typically at the DFT level, the potential energy surface for a single guest molecule can be obtained, and the combined with guest--guest classical interatomic potentials in a series of GCMC simulations. This approach has been demonstrated by Chen et al. on the study of methane adsorption on HKUST-1 (also known as Cu\e{3}(btc)\e{2}), a MOF with coordinatively unsaturated metal sites.\cite{Chen2011} This method combines a \textbf{DFT level of description of the host--guest interactions} with a full sampling of the phase space of the system, thus accounting for entropy. It is, however, only possible for atomic guests (including rare gases) or small spherical molecules (such as methane): for nonspherical molecules, the additional rotational degrees of freedom make the approach computationally prohibitive.

\subsection{Adsorption in flexible MOFs\label{sec:flexible}}

Although Monte Carlo simulations in the Grand Canonical ensemble, as presented at the beginning of this section, are considered the gold standard in the simulation of adsorption in nanoporous materials, they rely on a very strong assumption: that the host material is rigid. This approximation is very reasonable when it comes to the thermodynamics of adsorption in most porous inorganic materials, such as zeolites, where framework flexibility is limited. However, because metal--organic frameworks are based on weaker bonds and interactions (coordinative bonds, $\pi$--$\pi$ stacking, hydrogen bonds, etc.), that are responsible for their intrinsic structural flexibility. The organic--inorganic connections therefore allow underconstrained structural linkages that are responsible for soft mechanical properties, and organic linkers with side chains allow for local dynamics of the host framework. This flexibility of MOFs can be triggered upon adsorption, leading to large-scale structural changes in the MOF framework and subsequent alteration of its physical and chemical properties, including adsorption itself. Such flexible MOFs, sometimes called Soft Porous Crystals,\cite{Kitagawa_SPC} are in growing number.\cite{Lin2014, Schneemann2014, Coudert2015} There is therefore a need for molecular simulations methods describing the interplay between adsorption and host flexibility, going beyond the GCMC method. When the flexibility occurs in the form of a clear transition between well-defined and identified crystallographic structures, a simple ``two state'' solution can be used, studying the adsorption in both rigid structures through GCMC.\cite{FairenJimenez2011, Ania2012} However, more complex systems require a more direct approach, accounting directly for their flexibility in molecular simulations.

From a thermodynamics point of view, the adsorption of molecular fluids inside deformable hosts is most appropriately described in the \textbf{osmotic thermodynamic ensemble}, where the control parameters are the number of molecules of the host framework $N\e{host}$, the chemical potential of the adsorbed fluid $\mu\e{ads}$, the mechanical constraint $\sigma$ exerted on the system and the temperature $T$.\cite{Coudert2008} Direct molecular Monte Carlo simulations of adsorption in this open ensemble are possible, where the Monte Carlo moves typically used in GCMC are supplemented with ``volume change'' moves (see Figure~\ref{fig:MC}). However, the efficiency of these simulations are greatly improved by performing simulations in a hybrid MC/MD setup (also called Hybrid Monte Carlo, or HMC).\cite{Duane1987, Chempath2003, RASPA} In HMC, short molecular dynamics trajectories are considered as Monte Carlo moves, allowing to better sample the host framework's flexibility by following its collective motions.

\begin{figure*}[t]\centering
\includegraphics[width=0.9\textwidth]{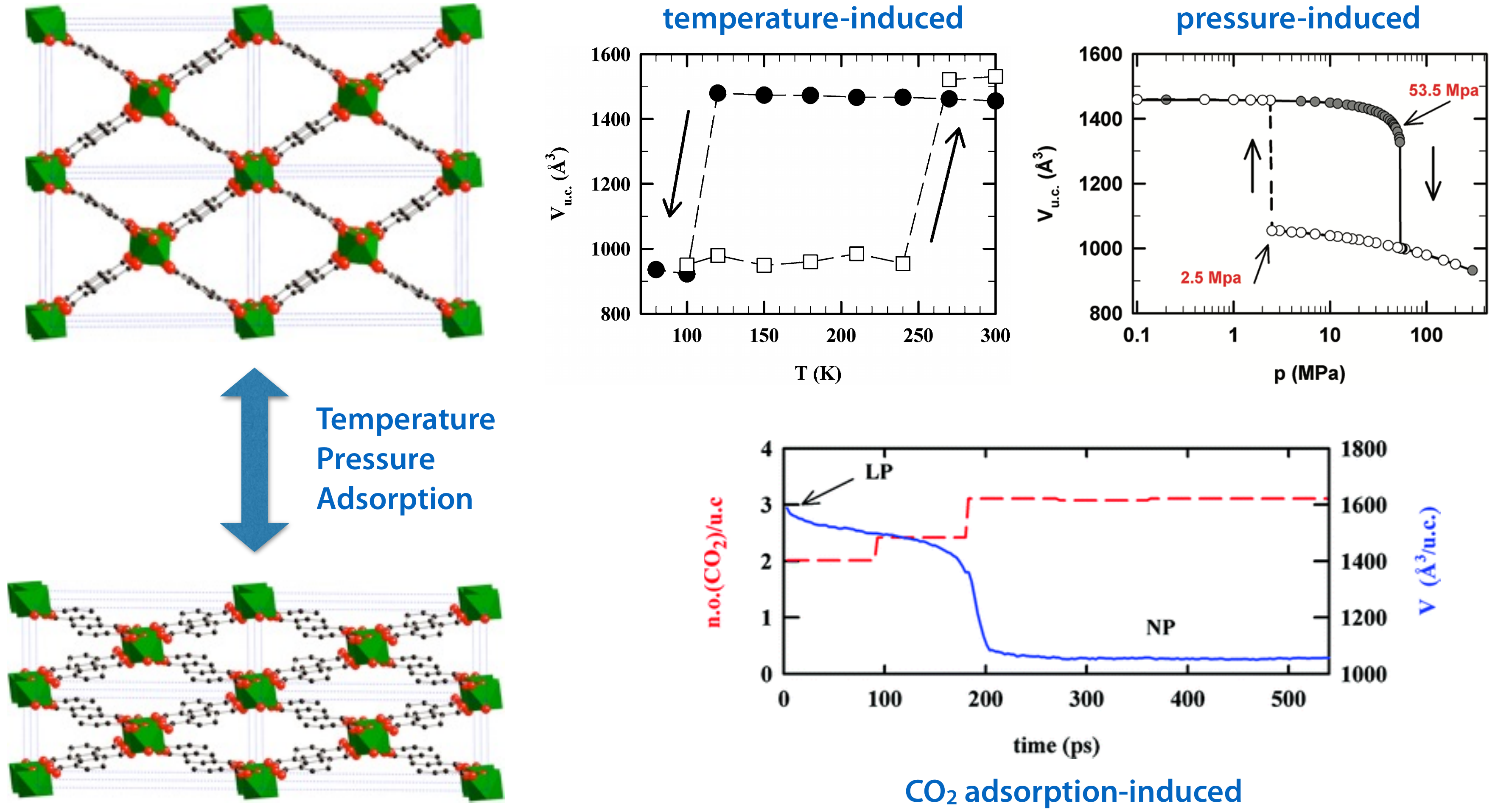}
\caption{\label{fig:Ghoufi}Left: the two metastable phases of ``breathing MOF'' MIL-53(Cr): large-pore (top) and narrow-pore (down). Right: structural transitions observed by Hybrid Monte Carlo molecular simulations, under stimulation by temperature, mechanical pressure, and carbon dioxide adsorption. Adapted with permission from refs.~\citep{Ghoufi2012} and \citep{Ghoufi2010b}. Copyright 2010, 2012 American Chemical Society.}
\end{figure*}

Maybe one of the most striking example of osmotic Monte Carlo simulations is that reported by Ghoufi et al. on the CO\e{2} adsorption-induced breathing of MIL-53(Cr).\cite{Ghoufi2010b, Ghoufi2012} This study showed that HMC in the osmotic ensemble was able to describe both thermal, mechanical and adsorption-induced structural transitions between the two phases of MIL-53(Cr), based on an \emph{ad hoc} force field describing the material (as depicted in Figure~\ref{fig:Ghoufi}). This study also demonstrated the main issue with direct HMC simulation, namely the widely hysteretic nature of the transitions and the difficulties in overcoming free energy barriers associated with the transition and determining the thermodynamic equilibrium of the system.

Another way to perform molecular simulation in the osmotic ensemble, avoiding the convergence issues of the direct Hybrid Monte Carlo method, is to rely on free energy methods to calculate the osmotic potential of the system for each possible value of chemical potential. These methods require the use of non-Boltzmann sampling in (guest loading, volume) parameter space in order to fully characterize the adsorption thermodynamics as well as the material's response to adsorption. In addition, they provide information on both the thermodynamic equilibrium as well as all metastable states of the system. Three different variants have been proposed: via thermodynamic integration based on GCMC simulations as performed by Watanabe et al.,\cite{Watanabe2008, Sugiyama2012} through the Wang--Landau algorithm as done by Bousquet et al.,\cite{Bousquet2012, Bousquet2013} or similarly with the Transition-Matrix Monte Carlo sampling as proposed by Shen et al.\cite{Shen2014} All three methods were demonstrated on model systems, and although they appear promising they have not yet been applied to atomistically-detailed molecular frameworks.

Finally, a third way to model adsorption in Soft Porous Crystals is the use of thermodynamics-based analytical models (see Ref.~\citep{Cou2011} for a review on this topic). This is achieved by writing down the equations for adsorption in the osmotic ensemble and introducing simple approximate expressions for certain quantities, such as describing the adsorption isotherms can be modeled by classical equations such as Langmuir or Langmuir-Freundlich, which can be either fitted from experimental data\cite{Coudert2008, Wang2009} computed from GCMC simulations of rigid frameworks,\cite{Gee2013, Zhang2014, Numaguchi2014} or calculated through parameterized equations of state.\cite{Ghysels2013} This approach was first used to study stepped adsorption isotherms in bistable materials of the \emph{gate opening} and \emph{breathing} families,\cite{Coudert2008, Cou2009} and later extended to take into account the influence of temperature\cite{Bou2009} and mechanical pressure.\cite{Ghysels2013}

\begin{figure*}[t]\centering
\includegraphics[width=0.8\linewidth]{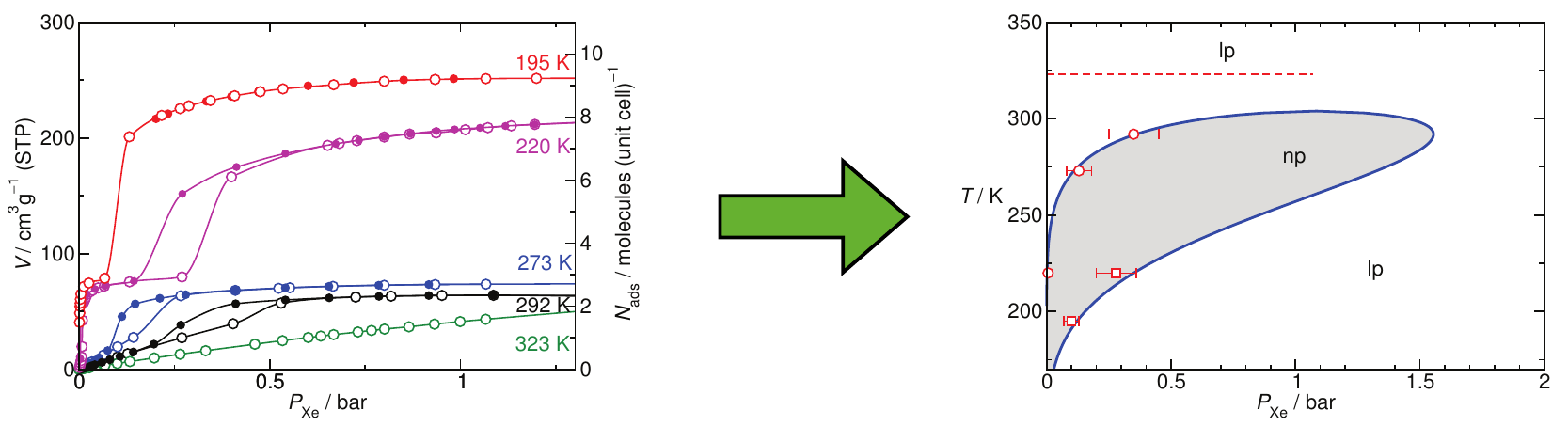}
\caption{\label{fig:phasediag}From a set of experimental adsorption isotherms in a flexible MOF, here for xenon in the ``breathing'' framework MIL-53(Al), analytical thermodynamic models allow the construction of a full (temperature, gas pressure) phase diagram. Reproduced from ref.~\citep{Bou2009}. Copyright 2009 Wiley-VCH Verlag GmbH \& Co. KGaA, Weinheim.}
\end{figure*}

Such analytical models not only help rationalized the behaviors observed experimentally, and shed light into their key thermodynamic factors and driving forces, but they can also have predictive value. This is, in particular, the case of the OFAST model\cite{Coudert2009, Coudert2010} (osmotic framework adsorption solution theory), extending the widely-used IAST coadsorption model to the osmotic ensemble. Based solely on experimental pure component adsorption isotherms, the OFAST model is able to predict coadsorption in flexible MOFs, and has been fully validated by comparison against experimental data, e.g. on the coadsorption of CO\e{2}/CH\e{4} mixtures in MIL-53(Al).\cite{Ortiz2012}

\subsection{Comparing simulations and experiments\label{sec:ads_exp}}

This short review on the computational modeling of adsorption in MOFs cannot be complete without a few words on the comparison between computational and experimental data --- and the pitfalls thereof. It is always desirable to validate a simulation methodology by comparing its results to available experimental data, when possible. However, the ample literature on MOF adsorption shows that simulation results and experimental data very often differ quantitatively\cite{Chowdhury2009} (as was already noted in the early work in the area\cite{Garberoglio2005}). Moreover, the differences can be in some cases rather large, without the simulation being necessarily at fault: there is also large variability in experimental measurements of adsorption data, and primarily adsorption isotherms. The possible causes for this variability are many:
\begin{enumerate}
	\item Dependence on the measurement technique used: volumetric vs. gravimetric, equilibration times, etc.\cite{Sing1999, SilvestreAlbero2009} Experimental isotherms may in some cases not truly represent the thermodynamic equilibrium, especially when adsorption kinetics is slow (e.g., bulky molecules in small pore MOFs, or occurrence of structural transitions).
	\item Impact of the activation procedure and dependence on the history of the sample. Residual amounts of solvent, organic linker or templates in the pores of a MOF can severely diminish its accessible pore volume and specific surface area.
	\item Influence of the textural properties, with key parameters such as crystal size distribution, nature of the external surfaces, crystal shape, etc. Adsorption on MOF samples of nanoscopic sizes occurs both inside the pores and on the external surface,\cite{Zhang2014} or at crystallographic line or plane defects. Molecular simulations almost always assume a bulk behavior (i.e. an infinite crystal), while crystal sizes in MOFs vary widely depending on material, synthesis conditions, etc.
	\item Presence of defects, in particular missing organic linkers in the crystalline structure. Depending on synthesis conditions, some MOFs can present large amounts of defects within their crystalline structure. These defects can have a dramatic impact on adsorption properties, typically increasing the pore volume and pore sizes,\cite{Sarkisov2011, Wu2013, Fang2014_JACS, Barin2014} but also affecting the host--guest interactions and the chemical nature of the internal surface of the material.\cite{Ghosh2014} Moreover, if the defects are organized rather than randomly dispersed throughout the MOF structure (e.g., correlated disorder in UiO-66\cite{Cliffe2014}), the effects on adsorption can be expected to be even more important.
\end{enumerate}

For all these reasons, it is highly recommended to perform a full characterization of the pore volume of MOFs, comparing their geometrical properties to the experimentally measured surface areas, pore volume, pore size distribution, etc.\cite{Walton2007, Duren2007} This is a necessary prerequisite to confirm that the simulations, performed on a perfect crystal structure, can match the experimental system and that any comparison of adsorption isotherms is meaningful. In some cases where the geometrical and experimental surface areas differ, it is possible to adjust the simulation results by scaling the adsorbed quantities by an empirical factor.\cite{Duren_COPS, Sillar2012} The reasoning behind this is to account for changes in pore volumes through blocked pores or missing linkers, which are probably the most common issues. The scaling factor is thus dependent on the material studied, the synthesis and activation conditions, the exact form and textural properties of the sample used experimentally\ldots Scaling factors used in the literature typically vary between 0.7 and 1. The use of such a scaling procedure allows one to compare or extrapolate computational data for different guests in the same adsorbent material. However, this ``quick fix'' cannot truly replace a better understanding of the origins of the discrepancy.

In conclusion, caution should be taken in comparing simulated adsorption results with experimental ones, and in particular when adjusting computational parameters (such as force field parameters) to fit certain experiments: one should do so not on a few isolated adsorption isotherms, but only based on a comprehensive experimental dataset (adsorption isotherms in wide temperature and pressure ranges, heats of adsorption, etc.).

\section{Perspectives}

Ending this introductory review of computational studies of metal--organic frameworks and their properties, we highlight some of the very recent work and remaining open questions that seem, from our perspective, to be important challenges for the development of the MOF field.

\subsection{Modeling of defects and disorder}

The occurrence of defects and their correlations have long been recognized to play a central role in the physical and chemical properties of materials. The study of defects in crystalline compounds forms an important part of solid-state chemistry and physics, as can be seen by the large body of literature concerning dense and porous inorganic materials.\cite{Lalena2010, auerbach2003handbook} Yet, the importance of the role of defects and disorder in metal--organic frameworks is only starting to emerge, and is still rarely studied in the existing literature, both experimental and theoretical. Yet, there is evidence that defects and disorder play an important role in the function of several existing MOFs. The poster child for this is the UiO-66(Zr or Hf) family of structures, of high interest because of their thermal, mechanical and chemical stability. Studies in the past three years have shown that UiO-66 materials can contain significant amount of missing-linker defects,\cite{Wu2013, Oien2014} which have a crucial impact on the adsorption and catalytic properties of the material.\cite{Vermoortele2013} The concentration of these defects can be tuned during synthesis;\cite{Wu2013} they do not occur randomly in the structure but are correlated and form nano-scaled domains of well-defined (\emph{reo}) topology.\cite{Cliffe2014} Moreover, the introduction of controlled heterogeneity in MOFs, without loss of its ordered structure, can lead to the creation of more complex structures and pore environments,\cite{Tu2014, Furukawa2015} introducing additional functions into known topologies and structures.\cite{Fang2014, Taylor2015} A recent review on the topic can be found in Ref.~\citep{Fang2015}.

In stark contrast to the few examples given above, most MOF structures reported in the literature feature a crystallographic structure and basic characterization of porosity, as well as some macroscopic study of their function (adsorption, catalytic activity, etc.). Because computational chemistry techniques are based on periodic representations of the crystalline structures, they model perfect materials with no defects and no disorder. A colleague summarized this first-order approach, tongue-in-cheek, as follows: for the most part, we are dealing with defects by \emph{``ignoring them, but invoking them as reason for any difference between modeling and experiments''}.\cite{Cramer_tweet}

Yet, some recent studies have used quantum chemistry and molecular simulation techniques to understand and predict the impact of defects on the properties of MOFs. At the quantum chemical level, a good example is the study by Chizallet et al. using DFT calculations of the catalysis of transesterification in ZIF-8, on acido-basic sites located at defects or at the external surface of the material.\cite{Chizallet2010} Another one is the influence of defect concentration on the structural, mechanical and thermal properties of UiO-66(Hf),\cite{Cliffe2014} giving microscopic insight into the occurrence of defect-dependent thermal densification and colossal negative thermal expansion (NTE) measured experimentally.\cite{Cliffe2015} At the classical level, Ghosh et al. have studied the impact of defects on adsorption properties of water in UiO-66(Zr) through Grand Canonical Monte Carlo simulations.\cite{Ghosh2014}   

Though there have been a few studies (listed above) on the impact of defects on MOF properties, it is worth noting that the bigger question has not been addressed so far from the theoretical point of view: \textbf{what causes certain MOFs to present defects, and when and why are these defects correlated?} There is experimental evidence that the occurrence of defects in MOFs can be random in some cases, and correlated in others. For example, structures obtained by linker-exchange\cite{Karagiaridi2012, Karagiaridi2014} show no correlation between the substituted linker positions, i.e. the resulting partially solvent-exchange MOF is a solid solution of the two parent MOFs, and features a random arrangement of linkers.\cite{Deria2014} On the other hand, missing-linker defects in UiO-66(Hf) exhibit correlated disorder, with the occurrence of nano-scaled domains of well-defined topology.\cite{Cliffe2014} Modeling studies have, so far, been unable to address this kind of very fundamental questions. Nor have there been any computational studies on noncrystalline disordered MOF structures, {i.e.} \textbf{molten, glassy, and amorphous MOFs}.\cite{Bennett2010, Besara2011, Bennett2014, Bennett_arxiv}

\subsection{Databases for high-throughput screening}

One of the areas that has seen fast-paced development in the last few years is that of high-throughput computational screening of porous materials in general, and metal--organic frameworks in particular. As described in some more detail in Section~\ref{sec:screening}, this is the result of the conjunction of a few factors: (i) a large number of know MOF structures, (ii) the availability of both computational power and advanced modeling techniques, and (iii) a political incentive to speed up the discovery of novel materials. This has lead to the development of two different types of large-scale databases of MOF structures: hypothetical MOF structures generated by combinatorial approaches (of the order of 100k structures);\cite{Wilmer2012} and ``computation-ready'' structures derived from experimental crystallographic data (of the order of 5k structures).\cite{Chung2014} The availability of these databases offers great opportunities for high-throughput computational screening of materials for specific applications, as has already been done for adsorption of hydrogen,\cite{Colon2014_JPCC, Gomez2014} methane,\cite{Wilmer2012, Simon2015_EES} carbon dioxide,\cite{Lin2012} noble gases,\cite{Sikora2012, Simon2015} etc. But this also raises some novel questions and challenges about how best to exploit these databases for materials discovery.

Focusing first on databases of hypothetical structures, it might be here interesting to draw some parallels with the existing databases of hypothetical zeolitic structures,\cite{Earl2006, Deem2009} which have existed for nearly ten years now and are of similar size (up to two million structures). There also, like for MOFs, the overwhelming majority of screening studies have focused on adsorption properties, relying both on computationally-cheap geometrical characterization (pore size distributions, pore diameters, surface area, pore volume; see Section~\ref{sec:geom}) and GCMC calculations on candidate structures selected on geometric criteria. This approach has yielded valuable insight into the relationship between geometric properties and adsorption performances,\cite{Wilmer2012} as well as the intrinsic limits of these materials.\cite{Simon2015_EES} It is, however, rather uncertain how this approach can be extended to the study of other properties, either for MOFs or zeolites: \textbf{simple and ``cheap'' descriptors are hard to identify for other functions} such as catalytic activity, thermal and mechanical properties, luminescence, flexibility, \ldots

Furthermore, given the large number of structures available, the use of a few descriptors and their correlation to key properties is typically an overdetermined problem. Even simply plotting the data generated can become difficult for multidimensional data sets: two-dimensional correlation plots, for example, are limited to testing hypotheses formulated \emph{a priori}. There is thus a real need to \textbf{use more sophisticated data mining tools} and to \textbf{move away from predetermined descriptors}. Some steps have been taken along this path in the use of quantitative structure--property relationship (QSPR) backed by machine learning algorithms, for example by Fernandez et al. to identify high-performing MOFs for carbon dioxide capture.\cite{Fernandez2014} Thornton et al. used a combination of molecular simulation and machine-learning techniques to identify candidate zeolites for catalytic reduction of carbon dioxide, from a sample of 300 thousand zeolite structures (hypothetical and experimental).\cite{Thornton2015} At a much smaller scale, {Y{\i}ld{\i}z} et al. have demonstrated the potential of Artificial Neural Networks (ANN) to predict adsorption properties on the case of hydrogen gas storage in thirteen MOFs with high surface areas.\cite{Yildiz2015} 

Another issue that needs to be addressed is that of the \textbf{experimental feasibility} of the hypothetical MOFs. Hypothetical zeolite databases have existed for quite some time, yet from their systematic exploration no real structure has been synthesized and tested for practical applications, and the question of why so few zeolites are observed while there are so many hypothetical frameworks remains open.\cite{Majda2008, Blatov2013, Li2013} Though MOFs seem to far a little better here, with several examples of materials synthesized after they have been selected by computational design,\cite{Farha2010} the important questions of experimental feasibility and \textbf{computational rational design} remain. Among the currently-listed hypothetical MOF structures, which are experimentally accessible to synthesis and present \textbf{sufficient mechanical, thermal and chemical stability for practical applications}? Can we use the tools of theoretical chemistry to help guide the synthesis of novel MOFs identified for their properties? Can we help predict the conditions (reagents, solvents, temperature, etc.) under which a given MOF may be obtained experimentally?

Finally, databases built from experimental structures, such as CoRE MOF,\cite{Chung2014} raise important questions as to the \textbf{curation of the database} and the \textbf{relevance of the ``cleaned-up'' structures to the properties of the original experimental materials}. The generation of computation-ready structures from crystallographic data involves an automated procedure to remove solvent and disordered guests. This procedure has severe limits. First, it assumes that all coordinated solvent molecules can be removed from the structure: this is true in some cases (water molecules bound to HKUST-1 metal centers, for example) but not in general. Indeed, there exist some MOFs with included solvent where the very stability of the framework is crucially dependent on the presence of the solvent molecules (e.g. water, methanol, etc.), and which cannot be activated without loss of crystallinity.\cite{Edgar2001, Tian2014, BouesselduBourg2014} Secondly, it assumes that there is no relaxation upon solvent removal, which is a rather crude approximation, especially in the case of soft porous crystals. Nonetheless, this large-scale database is the first of its kind and is likely to prove useful for understanding fundamental principles of MOFs and a basis for computational selection of structures for targeted applications.

\subsection{Stimuli-responsive MOFs}

With the large number of new MOF structures synthesized and characterized every year, one of the empirical patterns appearing is the common occurrence of flexibility of these framework materials. There is a rapidly increasing number of framework structures whose flexibility manifests in the form of large-scale structural transformations induced by external stimulation of physical or chemical nature: changes in temperature, mechanical constraints, guest adsorption, light exposure, etc. A number of different terms have been used to refer to this behavior, including \emph{smart materials}, \emph{soft porous crystals},\cite{Kitagawa_SPC} \emph{dynamic frameworks}, \emph{flexible frameworks}, and \emph{stimuli-responsive materials}.\cite{Coudert2015} The realization, over the past ten years, of the prevalence of these flexible materials and their potential for applications has lead to the emergence of an entire subfield of computational methods dedicated to their modeling. The existing literature on this topic has been described in Sections~\ref{sec:flexible} (adsorption--induced structural transitions) and Section~\ref{sec:mechanical} (mechanical properties). This is, however, an area in fast development and with many challenges still left open.

The first is the need for the \textbf{systematic development of high-accuracy force fields for highly flexible materials}. Force fields for MOFs with high degree of flexibility, such as MIL-53 or ZIFs, require a good accuracy in the description of the intramolecular interactions of the framework, and in particular the low-frequency phonons modes. For this, two categories of force fields have been used so far. The first, and dominant category, is the use of transferable intramolecular force fields for organic molecules (e.g., from the AMBER parameter database\cite{Cornell1995}) along with hand-tuned metal--organic bending and torsion potentials. The later are then optimized to reproduce known structural properties and vibration frequencies,\cite{Zheng2012, Hu2012, Wu2014} or experimental data such as adsorption isotherms and structural transitions.\cite{Salles2008, Coombes2009, Zhang2013} This approach is not entirely generalizable, and the quality of the resulting force field is only validated on limited experimental data and cannot be systematically improved, as the optimization problem is typically underconstrained (too many parameters fitted on relatively little data). On the other hand, force fields derived from \emph{ab initio} data (as described in Section~\ref{sec:abinitio}) show clear, but that methodology is rather hard to apply to flexible materials, especially when it comes to the intramolecular terms. Still, some recent progress has been made in this area, giving hope that generic methodologies for \emph{ab initio} force fields of flexible MOFs can be a reality in the future.\cite{Vanduyfhuys2015}

The second area of development we want to highlight here is the recent trend towards better insight into the microscopic mechanisms of stimuli-responsiveness, with particular focus on space-resolved, time-resolved, and \emph{in operando} experimental measurements. Those are crucial to provide a better fundamental understanding of the fundamental nature of the transformations triggered by complex stimuli and have practical consequences for the design of novel materials in working conditions. Yet, relatively little theoretical and computational effort has been spent on this, most probably due to the very difficult nature of the issue. The questions that need to be answered include: how do \textbf{stimuli-induced transformations occur and propagate at the scale of the crystal?} What is their \textbf{kinetics} and \textbf{dependence on the history} of the material? What determines the possible metastable states of the system? How do \textbf{crystal size, shape, and textural properties} affect their physical and chemical properties, and ultimately their responsivity?

Some of the recent studies, both experimental and theoretical, have started to address this issue. On the experimental side, one striking such example was the recent observation of an transient state in the adsorption-induced structural transition of pillared MOF Zn\e{2}(ndc)\e{2}(dabco),\cite{Dybtsev2004} by using synchrotron grazing incidence diffraction measurements to determine separately the structures of a crystal's bulk and surface.\cite{Kondo2014} Kondo et al. demonstrated that, upon adsorption of bulky slow-diffusing molecules, the MOF featured a heterostructure with a guest-induced sheared phase at the surface coexisting with an unperturbed MOF structure in the core of the crystal. On the theoretical point of view, we presented some time ago a multiscale physical mechanism and a stochastic model of breathing transitions,\cite{Triguero2011} which is to our knowledge still the only example of such ``upscaling'', trying to address computationally the dynamics of structural transitions at the level of the crystal itself. Based on a simple Hamiltonian that describes the physics of host--host and host--guest interactions, we showed how the behavior of unit cells is linked to the transition mechanism at the crystal level through three key physical parameters: the transition energy barrier, the cell-cell elastic coupling, and the system size.\cite{Triguero2012}

Along a similar line, but on the different topic of MOF crystal growth, Yoneya et al. have recently used coarse-grained molecular dynamics to investigate the MOF self-assembly and the influence of metal--ligand coordination in this process.\cite{Yoneya2015} This is a first step toward a better understanding of MOF synthesis and growth at the microscopic scale, a topic that has proven crucial but difficult to address so far, in MOFs as well as in other porous materials.\cite{Jin2010, Auerbach2014}

\section*{Acknowledgments}

We thank Anne Boutin for a long-standing and continuing collaboration on the fascinating topic of modeling of nanoporous materials. We thank Romain Gaillac and Thomas Manz for critical reading of the pre-print. FXC acknowledges PSL Research University's financial support (projects DEFORM and MECADS) and GENCI for high-performance computing CPU time allocations: although no calculations were performed in the writing of this review, a large part of the work reviewed here requires access of scientists in the field to large supercomputer centers.

\section*{Abbreviations}

\newcommand\abbr[2]{\noindent\begin{tabular}{p{18mm}l} \hfill #1: & #2 \end{tabular}\\}
\newcommand\abbrcont[1]{\noindent\begin{tabular}{p{18mm}l} & #1 \end{tabular}\\}

\abbr{AASBU}{Automated Assembly of}\abbrcont{Secondary Building Units}
\abbr{ANN}{Artificial Neural Network}
\abbr{AUA}{Anistropic United Atom}
\abbr{BET}{Brunauer--Emmett--Teller}
\abbr{CC}{Coupled Cluster}
\abbr{CHelpG}{CHarges from Electrostatic Potentials}\abbrcont{using a Grid-based method}
\abbr{CSD}{Cambridge Structural Database}
\abbr{COF}{Covalent Organic Framework}
\abbr{CoRE MOF}{Computation-Ready Experimental MOF}
\abbr{DFT}{Density Functional Theory}
\abbr{GCMC}{Grand Canonical Monte Carlo}
\abbr{HMC}{Hybrid Monte Carlo}
\abbr{HPC}{High-Performance Computing}
\abbr{IAST}{Ideal Adsorbed Solution Theory}
\abbr{IRMOF}{Iso-Reticular Metal--Organic Framework}
\abbr{MC}{Monte Carlo}
\abbr{MD}{Molecular Dynamics}
\abbr{MMSV}{Morse--Morse--spline--van der Waals}
\abbr{MOF}{Metal--Organic Framework}
\abbr{NLC}{Negative Linear Compressibility}
\abbr{NTE}{Negative Thermal Expansion}
\abbr{OFAST}{Osmotic Framework Adsorption}\abbrcont{Solution Theory}
\abbr{PSD}{Pore Size Distribution}
\abbr{QSPR}{Quantitative Structure--Property Relationship}
\abbr{RAST}{Real Adsorbed Solution Theory}
\abbr{RCSR}{Reticular Chemistry Structure Resource}
\abbr{RESP}{Restrained ElectroStatic Potential}
\abbr{SBU}{Secondary Building Unit}
\abbr{SPC}{Soft Porous Crystal}
\abbr{TDDFT}{Time-Dependent Density Functional Theory}
\abbr{UA}{United Atom}
\abbr{UFF}{Universal Force Field}
\abbr{VST}{Vacancy Solution Theory}
\abbr{ZAG}{Zinc Alkyl Gate}
\abbr{ZIF}{Zeolitic Imidazolate Framework}
\abbr{ZMOF}{Zeolite-like Metal--Organic Framework}

\section*{References}

\bibliographystyle{model1a-num-names}
\bibliography{biblio}

\end{document}